\documentclass[letterpaper,titlepage,11pt]{article}

\usepackage[a4paper]{geometry}
\usepackage{amssymb,amsmath,amsthm,amscd}
\usepackage{tensor}
\usepackage{mathtools}
\usepackage{enumerate}
\usepackage[english]{babel}
\usepackage{comment}

\usepackage{bm}
\usepackage[usenames,dvipsnames]{xcolor}

\newcommand{\RR}{\mathbb{R}}

\newcommand{\inv}{^{-1}}
\newcommand{\pd}{\partial}

\newcommand{\vphi}{\varphi}

\newcommand{\xm}{x^-}
\newcommand{\xp}{x^+}
\newcommand{\LL}{\mathcal{L}}

\usepackage[T1]{fontenc}
\usepackage[utf8]{inputenc}
\usepackage{lmodern}

\usepackage[colorlinks=true,urlcolor=blue,citecolor=magenta]{hyperref}
\usepackage{amssymb,amsmath,amsfonts}
\usepackage{epsfig}
\usepackage{graphicx}
\usepackage{epstopdf}
\usepackage{caption}
\usepackage{subcaption}
\usepackage{amscd}
\usepackage{stmaryrd}
\usepackage{amsthm}
\usepackage{latexsym}
\usepackage{amsbsy}
\usepackage[english]{babel}
\usepackage{psfrag}
\usepackage{tabularx}
\usepackage{cite}

\allowdisplaybreaks[1]

\def\clock{{\count0=\time
           \divide\count0 60
           \ifnum\count0<10 0\fi\the\count0
           \multiply\count0 -60 \advance\count0 \time
           :\ifnum\count0<10 0\fi \the\count0
         }}
\newcommand{\timestamp}{{\small\vbox{\hbox{\tt\jobname.tex}
\hbox{\the\day/\the\month/\the\year, \clock}}}}

\newcommand{\FB}{\overline{F}}
\newcommand{\FBp}{\FB^+}

\newcommand{\FBm}{\FB^-}

\def\mT{\mathcal{T}}
\def\mR{\mathcal{R}}
\newcommand{\mP}{\mathcal{P}}
\newcommand{\mJ}{\mathcal{J}}

\newcommand{\mD}{\mathcal{D}}

\newcommand{\mK}{\mathcal{K}}
\newcommand{\mS}{\mathcal{S}}

\newcommand{\mFB}{\overline{\mathcal{F}}}
\newcommand{\mFBp}{\mFB^+}

\newcommand{\mFBm}{\mFB^-}

\def\mL{\mathcal{L}}

\def\mM{\mathcal{M}}
\def\mN{\mathcal{N}}

\def\mD{\mathcal{D}}

\def\mP{\mathcal{P}}

\def\mF{\mathcal{F}}

\begin{document}

\numberwithin{equation}{section}

\begin{titlepage}
\rightline{\vbox{   \phantom{ghost} }}

 \vskip 1.8 cm
\begin{center}
{\Huge \bf \boldmath
Zooming in on AdS$_3$/CFT$_2$\\ near a BPS Bound}
\end{center}
\vskip .5cm

\title{}
\date{\today}
\author{Jelle Hartong, Yang Lei, Niels Obers, Gerben Oling}

\centerline{\large {{\bf Jelle Hartong$^1$, Yang Lei$^2$, Niels A. Obers$^3$, Gerben Oling$^1$}}}

\vskip 1.0cm

\begin{center}

\sl $^1$ Institute for Theoretical Physics and Delta Institute for Theoretical Physics,\\
University of Amsterdam, Science Park 904, 1098 XH Amsterdam, The Netherlands\\
\sl $^2$ CAS Key Laboratory of Theoretical Physics, Institute of Theoretical Physics,\\
Chinese Academy of Sciences, 100190 Beijing, P.R. China\\
\sl $^3$ The Niels Bohr Institute, Copenhagen University,\\
Blegdamsvej 17, DK-2100 Copenhagen \O , Denmark
\vskip 0.4cm

\end{center}

\vskip 1.3cm \centerline{\bf Abstract} \vskip 0.2cm \noindent

Any $(d+1)$-dimensional CFT with a $U(1)$ flavor symmetry, a BPS bound and an exactly marginal coupling admits a decoupling limit in which one zooms in on the spectrum close to the bound.
This limit is an In\"on\"u--Wigner contraction of $so(2,d+1)\oplus u(1)$ that leads to a relativistic algebra with a scaling generator but no conformal generators.
In 2D CFTs, Lorentz boosts are abelian and by adding a second $u(1)$ we find a contraction of two copies of $sl(2,\mathbb{R})\oplus u(1)$ to two copies of $P_2^c$, the 2-dimensional centrally extended Poincar\'e algebra.
We show that the bulk is described by a novel non-Lorentzian geometry that we refer to as pseudo-Newton--Cartan geometry.
Both the Chern--Simons action on $sl(2,\mathbb{R})\oplus u(1)$ and the entire phase space of asymptotically AdS$_3$ spacetimes are well-behaved in the corresponding limit if we fix the radial component for the $u(1)$ connection.
With this choice, the resulting Newton--Cartan foliation structure is now associated not with time, but with the emerging holographic direction.
Since the leaves of this foliation do not mix, the emergence of the holographic direction is much simpler than in AdS$_3$ holography.
Furthermore, we show that the asymptotic symmetry algebra of the limit theory consists of a left- and a right-moving warped Virasoro algebra.

\end{titlepage}

\newpage
\tableofcontents

\section{Introduction}

The idea of holography has become the most powerful tool in understanding theories of quantum gravity.
Its most celebrated realization is seen in the AdS/CFT correspondence  \cite{Maldacena:1997re,Witten:1998qj} which relates a general relativistic quantum gravity theory on an asymptotically anti-de Sitter (AdS) spacetime to a conformal field theory (CFT) living on the boundary of AdS.
While this duality has led to amazing progress in the last decades, with impact on a wide range of areas in theoretical physics, there are many fundamental questions that remain unanswered.
One widely studied route towards gaining deeper insight is by taking consistent limits of the correspondence, therewith simplifying both sides while still retaining non-trivial features.
Examples include the BMN limit \cite{Berenstein:2002jq}, the limit considered by Kruczenski \cite{Kruczenski:2003gt} as well as the closely related Spin Matrix Theory limit of Ref. \cite{Harmark:2014mpa}.
Other actively pursued directions are to consider
  i). non-AdS vacua within Einstein gravity \cite{Balasubramanian:2008dm,Son:2008ye,Taylor:2015glc} or
  ii). bulk theories that are Chern--Simons, higher spin or non-Einsteinian gravity theories like Ho\v rava--Lifshitz gravity\footnote{
For the discussion in this paper it is relevant to note that Ho\v rava--Lifshitz gravity \cite{Horava:2009uw} can be reformulated as a theory of dynamical Newton--Cartan geometry \cite{Hartong:2015zia}, and, moreover in three dimensions these theories  and generalizations thereof can be formulated \cite{Hartong:2016yrf} (see also \cite{Papageorgiou:2009zc,Bergshoeff:2016lwr}) as  Chern--Simons theories on non-Lorentzian kinematical algebras, such as the Bargmann or Newton--Hooke algebra.
Chern--Simons theories also play a role in another avatar of non-AdS holography \cite{Hofman:2014loa} involving warped CFTs (see for example \cite{Hofman:2011zj,Detournay:2012pc,ElShowk:2011cm,Song:2011sr,Song:2017czq,Jensen:2017tnb}) as boundary field theories.
Torsional Newton--Cartan geometry was first observed in the context of non-AdS holography in Refs. \cite{Christensen:2013lma,Christensen:2013rfa,Hartong:2015wxa}.
It is worth emphasizing that non-AdS holography is a vast subject, including for example also non-UV conformal models (see \cite{McGreevy:2009xe,Gursoy:2016ebw} for recent reviews).
However, for the purposes of this paper, we focus on the subset of theories that have nonrelativistic symmetries}.

In this paper we will study a decoupling limit of the AdS$_3$/CFT$_2$ correspondence that can be formulated in the bulk as a Chern--Simons theory that is not equivalent to Einstein gravity.
On the field theory side, the limit utilizes a $U(1)$ flavor symmetry, for example an R-symmetry of a superconformal field theory, and zooms in on the spectrum close to the lightest charged state of the theory on the cylinder.
Thus we are zooming in on the sector of the theory near a BPS bound, much in the spirit of the Spin Matrix Theory proposal of \cite{Harmark:2014mpa} which considers limits to critical points of ${\cal{N}}=4$ supersymmetric Yang--Mills theory, keeping only one-loop corrections to the BPS states.
On the gravity side, the limit results in a Chern--Simons (CS) action that is defined on an algebra that can be viewed as a different real form of the complexified versions of the algebras used in the Chern--Simons (Newton--Cartan) gravity theories of \cite{Hartong:2016yrf,Papageorgiou:2009zc,Bergshoeff:2016lwr}.
In particular, our CS theory describes a novel version of non-Lorentzian geometry, which we call pseudo-Newton--Cartan (pseudo-NC) geometry.
These theories have the feature that the holographic direction emerges  in a much simpler way than in AdS because there is a special foliation structure associated with it.

While our primary focus will be on AdS$_3$/CFT$_2$, the limit we consider is quite general, as any $d+1$ dimensional CFT with a $U(1)$ flavor symmetry, a BPS bound and a free coupling constant admits a limit in which one zooms in on the spectrum close to the lightest charged state of the theory on $\mathbb{R}\times S^d$.
In further detail, after using the state-operator map, the BPS operators have a conformal weight that is equal to the energy of the lightest charged state in units of the sphere radius.
Using the BPS condition this energy in turn is equal to a $U(1)$ charge $Q$.
We assume that the theory has a free marginal coupling constant $g$ that can be used to compute the one-loop corrections to the conformal dimension away from the BPS bound.
By turning on a `chemical potential'\footnote{
We write `chemical potential' in quotation marks since this is not a usual chemical potential corresponding to a time component of a background $U(1)$ gauge field.
Rather, it is the radial component of a background $U(1)$ potential in radial quantization.
} for the charge $Q$
we can offset the dilatation operator $D$ to $D-Q$ which still has a nonnegative spectrum.
This new dilatation operator has order $g$ corrections when we turn on the interactions perturbatively. The limit zooms in on this 1-loop piece of the dilatation operator. The symmetry algebra in this limit is an In\"on\"u--Wigner contraction of $so(2,d+1)\oplus u(1)$ that leads to a relativistic algebra with scale but no conformal generators.

For the specific case of 2D CFTs the Lorentz generator is abelian and it turns out to be useful to add a second $u(1)$.
One then considers an In\"on\"u--Wigner contraction of $so(2,2)\oplus u(1) \oplus u(1)$, i.e. two copies of $sl(2,\mathbb{R})\oplus u(1)$.
The resulting contracted algebra is then two copies of $P_2^c$, the two-dimensional centrally extended Poincar\'e algebra.
This algebra admits an infinite-dimensional extension, namely a left- and right-moving warped Virasoro algebra, which turns out to appear as the asymptotic symmetry of our novel bulk gravity dual.
We show that the CS action on two copies of $P_2^c$ can be obtained by applying our contraction to the CS action on two copies of $sl(2,\mathbb{R})\oplus u(1)$.
We then demonstrate that the entire phase space of asymptotically AdS$_3$ solutions of the $so(2,2)\oplus u(1) \oplus u(1)$ Chern--Simons theory can be mapped to the phase space of the limit theory.
This procedure requires a fixed radial chemical potential for one of the two $u(1)$ connections which is related to the background `chemical potential' needed to offset the dilatation operator close to a BPS bound.

We emphasize that since we can reach the full phase space of the $P_2^c$ theory from the
 $sl(2,\mathbb{R})\oplus u(1)$
theory, we can study many well-understood aspects of the AdS$_3$/CFT$_2$  correspondence in this limit.
For example, this procedure allows us to find the vacuum of the theory.
Following the coset procedure for homogeneous nonrelativistic spacetimes proposed in
\cite{Grosvenor:2017dfs} we show that this three-dimensional vacuum geometry corresponds to the
coset $ (\mathbf{P}_2^c \times \mathbf{P}_2^c)/(\mathbf{P}_2^c \times  U(1))$, where $\mathbf{P}^c_2$ stands for centrally extended 2D Poincar\'e  group.
We demonstrate that for this background 6 of the 8 generators of the $P_2^c\oplus P_2^c$ algebra are realized in terms of Killing vectors while the two central extensions are visible only once we study matter fields on these backgrounds.

Our limiting procedure of contracting $so(2,2)\oplus u(1)\oplus u(1)$ by turning on a radial chemical potential to zoom in on the lightest charged state of the theory on the cylinder bears a strong resemblance to viewing nonrelativistic limits as contractions of Poincar\'e$\oplus u(1)$.
There, one turns on a chemical potential to offset the Hamiltonian of the relativistic theory (splitting off the rest mass) before sending the speed of light to infinity \cite{Jensen:2014wha,Bergshoeff:2015uaa}.
In \cite{Bergshoeff:2016lwr} a contraction was studied along these lines of a CS theory on the relativistic algebra $iso(2,1)\oplus u(1)\oplus u(1)$. In the large speed of light limit the two $u(1)$ gauge fields correspond to two quantum numbers of a nonrelativistic particle, the rest mass and rest spin \cite{Papageorgiou:2009zc}.
This limit can be generalized in the presence of a cosmological constant leading to CS theory on what is known as the extended Newton--Hooke algebra.
For a negative cosmological constant this algebra is isomorphic to $E_2^c\oplus E_2^c$, with $ E_2^c$ the two-dimensional centrally extended algebra of the Euclidean two-plane.
The CS action on the extended Newton--Hooke algebra was considered in \cite{Hartong:2016yrf,Papageorgiou:2010ud}.
The limit of $so(2,2)\oplus u(1)\oplus u(1)$ we take in this paper leads to the algebra $P_2^c\oplus P_2^c$.
This can be viewed as a different real form of a complexified version of the $E_2^c\oplus E_2^c$ algebra.

We know from the CS theory on $E_2^c\oplus E_2^c$ that the geometry is described by Newton--Cartan geometry in which time plays a special role.
Essentially Newton--Cartan geometry is a covariant description of a special foliation structure where each leaf corresponds to a certain instant of time.
In the case of the CS theory on $P_2^c\oplus P_2^c$, the bulk geometry is a version of Newton--Cartan (NC) geometry that we call pseudo-Newton--Cartan geometry.
In this case the leaves are those corresponding to constant values of the emerging holographic coordinate and each such hypersurface has a two-dimensional Lorentzian signature.
This suggests that pseudo-Newton–Cartan geometry in the bulk provides, in some sense, a much simpler realization of the holographic paradigm than the one that employs Riemannian geometry.
Since the limit can be generalized to higher dimensions on the level of the algebra, we can speculate that the corresponding geometry (obtained by gauging the algebra) will appear more generally as a bulk dual in any dimension when zooming in on the spectrum close to the BPS bound.
Just like NC geometry can be obtained by gauging the Bargmann algebra  \cite{Hartong:2016yrf,Andringa:2010it,Bergshoeff:2014uea} in its usual real form, pseudo-NC geometry can be obtained by gauging a different real form of the complexified Bargmann algebra that we discuss here.

It is interesting to put the holographic correspondence that we obtain in this paper in the larger context of non-AdS holography.
So far, we have encountered roughly three classes of dualities.
In its original form, the AdS/CFT correspondence relates a relativistic bulk gravity to a corresponding dual (conformal) relativistic field theory living on the boundary.
For non-AdS holography, using for example asymptotically Schr\"odinger or Lifshitz spacetimes, there are setups with relativistic theories in the bulk,
e.g. Einstein--Maxwell-dilaton or Einstein--Proca-dilaton theories, and nonrelativistic field theories living on the boundary.
These boundary theories naturally couple to nonrelativistic geometries \cite{Son:2013rqa,Jensen:2014aia,Hartong:2014pma}, such as (torsional) NC geometry, as first shown in the holographic context in \cite{Christensen:2013lma,Christensen:2013rfa,Hartong:2015wxa,Hartong:2014oma}.
Additionally, it has been suggested that the latter field theories have
perhaps a more natural holographic realization with nonrelativistic gravity theories in the bulk \cite{Hartong:2016yrf,Hofman:2014loa,Kachru:2008yh,Griffin:2012qx}.
What we see here, somewhat unexpectedly, is that there is a fourth situation in which one has a nonrelativistic bulk gravity theory but a scale invariant relativistic field theory on the boundary.

It is natural to wonder how these scale invariant two-dimensional field theories that appear in our novel holographic correspondence relate to standard 2D CFTs.
Local unitary 2D scale invariant relativistic field theories for which the dilatation operator is diagonalizable and has a non-negative discrete spectrum admit currents for special conformal generators \cite{Polchinski:1987dy}, i.e. the charge algebra of symmetries of the theory is $sl(2,\mathbb{R})\oplus sl(2,\mathbb{R})$ which enhances to two commuting Virasoro algebras.
This infinite dimensional symmetry appears as the asymptotic symmetry algebra
in AdS$_3$ gravity, as shown in the seminal paper \cite{Brown:1986nw}.
In our case we will show that likewise the theory has a symmetry algebra that contains two copies of a warped Virasoro algebra and so has the symmetries of a CFT\footnote{
See also \cite{Guica:2017lia} for recent related work.
Note that these theories are different from the chiral warped CFTs studied in \cite{Hofman:2011zj}, since we have two copies of a warped Virasoro algebra.}.
To understand the nature of the two-dimensional field theory better one would have to work out the unitary irreducible representations of the two copies of the warped Virasoro groups.
We leave a more detailed analysis of this for the future and comment on related aspects in the discussion.

\subsection*{Outline and brief summary}
This paper is organized as follows.
In Section \ref{sec:CSalgebra} we present the various algebras
that play a role in our holographic construction and also present the CS action that appears in our limit.
In particular, in Section \ref{ssec:contractalgebra} we discuss the
In\"on\"u--Wigner contraction of $so(2,2)\oplus u(1) \oplus u(1)$ and show how this gives $P_2^c\oplus P_2^c$.
We also present its alternative form in terms of the three-dimensional extended Newton--Hooke algebra with a non-standard real form.
We then perform in Section \ref{ssec:contractCSaction} the same contraction at the level of the CS theory, resulting in a CS action
for two copies of $P_2^c$, which inherits a chiral structure from its relativistic parent action.
This action thus describes pseudo-Newton--Cartan gravity in three dimensions.
To put our limit in a broader perspective, we also discuss in Section \ref{ssec:higherdim}
the In\"on\"u--Wigner contraction of $so(2,d+1)\oplus u(1)$.
From the boundary point of view,
this results in a novel algebra that contains relativistic and scale symmetries, but no conformal symmetries.
From the bulk perspective we show that it corresponds to a different real form of the complexified $(d+2)$-dimensional
Newton--Hooke algebra.

We then turn in Section \ref{sec:phasespace} to studying the phase space of our limit theory by relating
it to the phase space of the parent theory.
To this end, we first show in Section \ref{ssection:metricdatap2c} how we can
obtain the most general  $P_2^c$ connection as a limit of the most general $sl(2,\mathbb{R})\oplus u(1)$
connection.
We then continue to show how the limit of the phase space of the
$so(2,2)\oplus u(1) \oplus u(1)$ CS theory acts in terms of metric data.
In Section \ref{ssec:vacuumsymmetry} we examine what happens to the Poincar\'e and global AdS${}_3$ geometries after the limit and study
the symmetries of the resulting vacua.
Finally, in Section \ref{ssec:bulk-scalar} we consider a bulk scalar field model where all symmetries, including central extensions, are explicitly realized.

In Section \ref{sec:asymptalgebra}, we then study the asymptotic symmetry algebra after the limit.
The main result is presented in Section \ref{ssec:infinitealphaASG} where we show that in the limit theory one obtains a chiral and antichiral warped Virasoro algebra.
To elucidate the appearance of this particular infinite-dimensional algebra, we repeat the same procedure in Section \ref{ssec:finitealphaASG} at finite value of the contraction parameter and obtain a more general form of the algebra in that case.
We then discuss in Section \ref{ssec:sectionuncouple} how this more general result relates to the naively expected asymptotic symmetry  algebra consisting of an uncoupled Virasoro and affine $u(1)$.
We end with a discussion and outlook in Section \ref{sec:discussions}.

Two appendices are included. Appendix \ref{app:so22-cs-review} reviews a number of useful elements
of $so(2,2)$ CS theory that we will draw on in the main parts of the paper, while
Appendix  \ref{sec:CSasymptsym} gives a detailed derivation of the asymptotic symmetries that one obtains from $sl(2,\mathbb{R})$ CS theory.
This should aid the reader in following our
derivations of the asymptotic symmetry algebra for the CS theory on $P_2^c\oplus P_2^c$.

\section{\boldmath Near-BPS limit of AdS Chern--Simons theory}
\label{sec:CSalgebra}
Three-dimensional Einstein gravity is simple due to the absence of local degrees of freedom.
On the other hand, it is also rich in boundary symmetries \cite{Brown:1986nw}.
Thus, it provides a useful arena to study holographic correspondences.
Furthermore, three-dimensional Einstein gravity can be formulated as a Chern--Simons (CS) theory \cite{Achucarro:1987vz,Witten:1988hc} which simplifies its boundary analysis.\footnote{
  Much work has been done to make this identification more precise.
  In particular, subtleties arise at the quantum level, see for example \cite{Witten:2007kt,Maloney:2007ud,Maloney:2015ina}, but we will not touch upon these issues here.
}
The Lagrangian of three-dimensional CS theory is
\begin{equation}\label{eq:chern-simons-action}
  \mathcal{L}_\text{CS}
    = \left\langle
        \mathbf{A}, d\mathbf{A}
          - \frac{2}{3}i\mathbf{A} \wedge\mathbf{A}
      \right\rangle.
\end{equation}
Here, the connection $\mathbf{A}$ is a Lie algebra-valued one-form.
The bracket denotes an invariant bilinear form on the algebra.
For a given basis $T_A$ of Hermitian generators of the Lie algebra, we denote its components by $\Omega_{AB}=\langle T_A, T_B \rangle$.
The structure constants of the algebra are defined by $[T_A, T_B] = i \tensor{f}{_{AB}^C}T_C$.
Invariance of the bilinear form corresponds to $\langle T_A, [T_B, T_C]\rangle = \langle [T_A, T_B], T_C \rangle$.

Three-dimensional Einstein gravity with negative cosmological constant corresponds to CS theory with gauge algebra $so(2,2)$.
This algebra splits in two simple $sl(2,\RR)$ factors.
Each of these factors has a unique invariant bilinear form given by its Killing form, which is fixed up to an overall constant.
See Appendix \ref{app:so22-cs-review} for further conventions and a detailed review of $so(2,2)$ CS theory.

In this paper, we will be concerned with a non-semisimple algebra.
In that case, there can be multiple parameters characterizing the invariant bilinear form.
However, our algebras will be such that $\Omega_{AB}$ can always be chosen to be nondegenerate and we will subsequently adopt this choice.
This means that all components of the connection enter in \eqref{eq:chern-simons-action}.
Some earlier examples of Chern--Simons theory with non-semisimple algebras include \cite{Papageorgiou:2009zc,Nappi:1993ie}, see \cite{Hartong:2016yrf,Bergshoeff:2016lwr} for more recent work.

\subsection{Contraction of 2D conformal algebra with abelian charges}
\label{ssec:contractalgebra}
From the boundary perspective, our starting point is a two-dimensional conformal field theory with two $u(1)$ flavor symmetries.
We are interested in an In\"on\"u--Wigner contraction of the global symmetry algebra $so(2,2)\oplus u(1)\oplus u(1)$ that zooms in on the state with lowest charge under the flavor symmetries.\footnote{
  Note that our initial algebra $so(2,2)\oplus u(1) \oplus u(1)$ is the bosonic part of the $\mN=(2,2)$ superconformal algebra, where the abelian currents correspond to the $R$-charge current in each chiral sector.
}
As we will see, the resulting algebra has a nondegenerate bilinear form, and can therefore be used to construct a Chern--Simons theory.
After identifying the correct limit of the algebra in the following, we will proceed to studying the contraction of the corresponding bulk gravity.

\subsection*{Relativistic conformal algebra}
In AdS$_3$/CFT$_2$, the $so(2,2)$ symmetry algebra has two interpretations.
First of all, it is the global conformal algebra of the conformal field theory living on the boundary of AdS$_3$.
In this incarnation, we will exhibit it using the standard basis of translations, boosts, dilatation and special conformal transformations, which satisfy
\begin{equation}\label{eq:confalgebra}
  \begin{split}
    &[P_a,K_b]=-2iD\eta_{ab}-2iM\epsilon_{ab}\,,\quad [D,P_a]=iP_a\,,\quad [D,K_a]=-iK_a\,,\\
    &[M,P_a]=i\epsilon_a{}^bP_b\,,\quad [M,K_a]=i\epsilon_a{}^bK_b\,.
\end{split}
\end{equation}
Here and in the following, we will use a lowercase Latin index $a=(0,1)$ for boundary components, so $\eta_{ab}$ and $\epsilon_{ab}$ are the 2-dimensional Minkowski metric and Levi--Civita symbol, respectively.
Boundary indices are raised and lowered with $\eta_{ab}$ and we set $\epsilon_{01}=+1$.

From a bulk gravity perspective, $so(2,2)$ is also the isometry algebra of AdS$_3$.
For this purpose, the natural generators are the bulk translations $T_A$ and bulk rotations $J_A$ given by
\begin{equation}\label{eq:AdSspacegens}
T_a=\frac{1}{2l}\epsilon_a{}^b(P_b+K_b)\,,\qquad T_2=\frac{1}{l}D\,,\qquad J_a=\frac{1}{2}(P_a-K_a)\,,\qquad J_2=M\,.
\end{equation}
They satisfy the following commutation relations,
\begin{equation}\label{eq:app-so(2,2)-alg}
[T_A,T_B] = \frac{i}{l^2} \tensor{\epsilon}{_{AB}^C} J_C, \quad
[J_A, J_B] = i\tensor{\epsilon}{_{AB}^C} J_C, \quad
[J_A, T_B] =i \tensor{\epsilon}{_{AB}^C} T_C,
\end{equation}
where $A=(a,2)=(0,1,2)$.
The parameter $l$ is the radius of curvature.
The bulk tangent space metric $\eta_{AB}=\text{diag}(-1,1,1)$ is used to raise and lower bulk indices and we set $\epsilon_{012}=+1$.

Finally, the $so(2,2)$ generators can be split into two chiral copies of $sl(2,\RR)$,
\begin{equation}
\label{eq:sl2-sl2bar-commutators}
[L_n,L_m]=i(n-m)L_{n+m}\,,\qquad [\bar L_n,\bar L_m]=i (n-m)\bar L_{n+m}\,.
\end{equation}
The left- and right-moving generators $L_m$ and $\bar L_m$ are given by
\begin{gather}
  L_{-1} = \frac{1}{2}\left(P_1+P_0\right), \quad
  L_0 = \frac{1}{2}\left(D+M\right), \quad
  L_1 = \frac{1}{2}\left(K_1-K_0\right),
  \\
  \bar L_{-1} =\frac{1}{2}\left(P_1-P_0\right), \quad
  \bar L_0 = \frac{1}{2}\left(D-M\right), \quad
  \bar L_1 = \frac{1}{2}\left(K_1+K_0\right)\,.
\end{gather}
Upon Wick rotation ($t=it_E$) the left- and right-moving coordinates $(\xp,\xm)$ become the complex coordinates $(z,\bar z)$.
Additionally, we assume that the 2D CFT has two flavor $u(1)$ symmetries generated by $Q_1$ and $Q_2$.
For future convenience, we define the following combinations of $u(1)$ generators,
\begin{equation}
\label{eq:N0-bar-N0-definition}
N_0 := \frac{1}{2}(lQ_1+Q_2),
\quad
\bar{N}_0 := \frac{1}{2}(lQ_1-Q_2).
\end{equation}
We will see
in the following
that these combinations naturally appear in the chiral decomposition of the contracted algebra.
We parametrize the invariant bilinear form on the left- and right-moving $sl(2,\mathbb{R})\oplus u(1)$ by
\begin{align}\label{eq:metricsl2+U1}
  2\langle L_0,L_0\rangle
    &=-\langle L_{-1},L_1\rangle
    =\gamma_s\,,
  \qquad
  \langle N_0,N_0\rangle=\frac{1}{2}\gamma_u\,,\\
  \label{eq:metricsl2+U1-bar}
  2\langle\bar L_0,\bar L_0\rangle
    &=-\langle\bar L_{-1},\bar L_1\rangle
    =-\bar\gamma_s\,,
  \qquad
  \langle\bar N_0,\bar N_0\rangle=-\frac{1}{2}\bar\gamma_u\,.
\end{align}

Let $\Phi$ be a field that transforms in a representation of the total symmetry group $so(2,2)\oplus u(1) \oplus u(1)$.
At the level of the algebra, a representation is given by
\begin{eqnarray}
  \label{eq:conformal-algebra-field-representation}
P_a\Phi & = & -i\partial_a\Phi\,,\nonumber\\
M\Phi & = & -i\left(\epsilon_a{}^bx^a\partial_b+s\right)\Phi\,,\nonumber\\
D\Phi & = & -i\left(x^a\partial_a+\Delta\right)\Phi\,,\\
K_a\Phi & = & -i\left(2\eta_{ac}x^cx^b\partial_b-x^2\partial_a+2\eta_{ab}x^b\Delta-2\epsilon_{ab}x^bs\right)\Phi\,,\nonumber\\
Q_1\Phi & = & q_1\Phi\,,\nonumber\\
Q_2\Phi & = & q_2\Phi\,.\nonumber
\end{eqnarray}
The field $\Phi$ thus carries the labels $\Delta$ (conformal dimension), $s$ (spin) and $q_1$, $q_2$ (charges).

\subsection*{Contracted algebra}
There exists a general procedure for obtaining nonrelativistic algebras from a contraction of relativistic ones, which is loosely referred to as In\"on\"u-Wigner contraction (see for example \cite{Bergshoeff:2015uaa} for a review).
In fact, this procedure does not exclusively yield nonrelativistic algebras, and has also been applied to studying the flat space limit of AdS holography \cite{Bagchi:2009my,Bagchi:2010eg,Gary:2014ppa}.

The procedure is as follows.
Define a basis for the initial algebra that depends on some parameter $\alpha$.
In this basis, the structure constants now depend on $\alpha$ but the algebra is still fundamentally unchanged.
However, by taking $\alpha \to \infty$ one obtains a \emph{contracted} algebra that is generically not isomorphic to the initial algebra.

Starting from $so(2,2)\oplus u(1) \oplus u(1)$, we want to consider the following contraction.\footnote{
  This is not the only possible contraction. For other options, see \cite{Basu:2017aqn}.
  }
Define the generators $\mathcal{P}_a$, $\mathcal{K}_a$, $\mathcal{D}$, $\mathcal{M}$, $\mathcal{N}$ and $\mathcal{S}$ by setting
\begin{eqnarray}
P_a & = & \alpha\mathcal{P}_a\,,\qquad K_a=\alpha\mathcal{K}_a\,,\nonumber\\
D & = & \frac{1}{2}\mathcal{D}+\alpha^2\mathcal{N}\,,
\qquad Q_1= - \frac{1}{2}\mathcal{D} + \alpha^2\mathcal{N}\,,\label{eq:genredef}\\
M & = & \frac{1}{2}\mathcal{M}+\alpha^2\mathcal{S}\,,
\qquad Q_2= - \frac{1}{2}\mathcal{M} + \alpha^2\mathcal{S}\,.\nonumber
\end{eqnarray}
From \eqref{eq:confalgebra} we find that they satisfy the following commutation relations,
\begin{equation}
\begin{split}
  &[\mP_a, \mK_b]
    = -2i \left(\frac{\mathcal{D}}{2\alpha^2} + \mathcal{N}\right)\eta_{ab}
      -2i \left(\frac{\mathcal{M}}{2\alpha^2} + \mathcal{S}\right)\epsilon_{ab},
    \\
  &[\mD, \mP_a]
    = i \mP_a,
    \qquad
  [\mD, \mK_a]
    = -i \mK_a,
    \\
  &[\mM, \mP_a]
    = i \tensor{\epsilon}{_a^b} \mP_b,
    \qquad
  [\mM, \mK_a]
    = i \tensor{\epsilon}{_a^b} \mK_b,
    \\
  &[\mN, \mP_a]
    = \frac{i\mP_a}{2\alpha^2},
    \qquad
  [\mN, \mP_a]
    = - \frac{i\mK_a}{2\alpha^2},
    \\
  &[\mS, \mP_a]
    = \frac{i \tensor{\epsilon}{_a^b} \mP_b}{2\alpha^2},
    \qquad
  [\mS, \mK_a]
    = \frac{i \tensor{\epsilon}{_a^b} \mK_b}{2\alpha^2}.
\end{split}
\end{equation}
At this point, we have only performed a basis transformation and the algebra is still unchanged.
However, by sending $\alpha\rightarrow\infty$, we obtain an \emph{inequivalent} algebra with
\begin{eqnarray}
&&[\mathcal{P}_a,\mathcal{K}_b]=-2i\mathcal{N}\eta_{ab}-2i\mathcal{S}\epsilon_{ab}\,,\nonumber\\
&& [\mathcal{D},\mathcal{P}_a]=i\mathcal{P}_a\,,\quad [\mathcal{D},\mathcal{K}_a]=-i\mathcal{K}_a\,,\label{eq:contractedalgebra}\\
&&[\mathcal{M},\mathcal{P}_a]=i\epsilon_a{}^b\mathcal{P}_b\,,\quad [\mathcal{M},\mathcal{K}_a]=i\epsilon_a{}^b\mathcal{K}_b\,.\nonumber
\end{eqnarray}
All other commutators vanish and one observes that $\mN$ and $\mS$ are now central elements.
This algebra, which we will call the \emph{scaling nonconformal algebra}, will be the central object of study in this paper.
The generators $\mathcal{M}$ and $\mathcal{P}_a$ form a 2D Poincar\'e subalgebra and $\mathcal{D}$ is a dilatation generator.
However, the $\mathcal{K}_a$ can no longer be thought of as conformal generators.
A representation of this algebra on a field $\Phi$ is given by
\begin{eqnarray}
\mathcal{P}_a\Phi & = & -i\partial_a\Phi\,,\nonumber\\
\mathcal{M}\Phi & = & -i\left(\epsilon_a{}^bx^a\partial_b+\sigma\right)\Phi\,,\nonumber\\
\mathcal{D}\Phi & = & -i\left(x^a\partial_a+\delta\right)\Phi\,,\\
\mathcal{K}_a\Phi & = & \left(2\eta_{ab}x^bN-2\epsilon_{ab}x^bS\right)\Phi\,,\nonumber\\
\mathcal{N}\Phi & = & N\Phi\,,\nonumber\\
\mathcal{S}\Phi & = & S\Phi\,.\nonumber
\end{eqnarray}

Like $so(2,2)$, the algebra \eqref{eq:contractedalgebra} can be split in two factors.
Here, the factors are given by a 2D Poincar\'e algebra with central extension, which we denote by $P_{2}^c$.
To see this we define
\begin{eqnarray}
\mathcal{L}_{-1} &=& \frac{1}{2}\left(\mathcal{P}_1+\mathcal{P}_0\right)
\,, \quad
\mathcal{L}_0 = \frac{1}{2}\left(\mathcal{D}+\mathcal{M}\right)
\,,\quad
\mathcal{N}_0 =\mathcal{N}+\mathcal{S}
\,,\quad
\mathcal{N}_{1} = \frac{1}{2}\left(\mathcal{K}_1-\mathcal{K}_0\right)
\,,
\\
\bar{\mathcal{L}}_{-1} &=& \frac{1}{2}\left(\mathcal{P}_1-\mathcal{P}_0\right),
\quad
\bar{\mathcal{L}}_0 =\frac{1}{2}\left(\mathcal{D}-\mathcal{M}\right)
\,,\quad
\bar{\mathcal{N}}_0 = \mathcal{N}-\mathcal{S}
\,,\quad
\bar{\mathcal{N}}_{1} = \frac{1}{2}\left(\mathcal{K}_1+\mathcal{K}_0\right)
\,.
\end{eqnarray}
The nonzero commutators of these generators are
\begin{equation}\label{eq:P11-finite-alpha}
\begin{gathered}
[\mathcal{L}_{-1}\,,\mathcal{L}_0]
= -i\mathcal{L}_{-1}\,,\qquad
[\mathcal{L}_{-1}\,,\mathcal{N}_1]
= -i\mathcal{N}_0 - \frac{i}{\alpha^2}\LL_0 \,,\qquad
[\mathcal{L}_0\,,\mathcal{N}_1]
=-i\mathcal{N}_1\,, \\
[\mN_0\,, \LL_{-1}] = \frac{i}{\alpha^2} \LL_-\,,\qquad
[\mN_0\,, \mN_{1}] = - \frac{i}{\alpha^2} \mN_{1}
\end{gathered}
\end{equation}
and likewise for the barred generators.
At finite $\alpha$, this is just a redefinition of $sl(2,\RR)\oplus u(1)$, but in the limit $\alpha\to\infty$ these are the commutation relations of $P_2^c$,
\begin{equation}\label{eq:P11}
[\mathcal{L}_{-1}\,,\mathcal{L}_0]
= -i\mathcal{L}_{-1}\,,\qquad
[\mathcal{L}_{-1}\,,\mathcal{N}_1]
= -i\mathcal{N}_0\,,\qquad
[\mathcal{L}_0\,,\mathcal{N}_1]
=-i\mathcal{N}_1\,.
\end{equation}
One can think of $\LL_{-1}$ and $\mN_1$ as translation generators in a two-dimensional Poincar\'e plane, with Lorentz boost $\LL_0$.
The central extension is given by $\mathcal{N}_0$.
This algebra can be viewed as a different real form of the complexified centrally extended 2D Euclidean algebra $E_2^c$.
The latter is sometimes referred to as the Nappi--Witten algebra \cite{Nappi:1993ie}.
We parametrize the most general invariant bilinear form on the two copies of $P_{2}^c$ using
\begin{eqnarray}
&&\langle\mathcal{L}_0, \mathcal{L}_0\rangle=\frac{1}{2}\gamma_1\,,\qquad\langle\mathcal{L}_0, \mathcal{N}_0\rangle=-\langle\mathcal{L}_{-1}, \mathcal{N}_1\rangle=\gamma_2\,, \label{eq:p2cmetric}\\
&&\langle\bar{\mathcal{L}}_0, \bar{\mathcal{L}}_0\rangle=-\frac{1}{2}\bar\gamma_1\,,\qquad\langle\bar{\mathcal{L}}_0,\bar{\mathcal{N}}_0\rangle=-\langle\bar{\mathcal{L}}_{-1}, \bar{\mathcal{N}}_1\rangle=-\bar\gamma_2\,.
\end{eqnarray}
We will later see that \eqref{eq:P11} has an infinite dimensional lift with generators $(\LL_m,\bar\LL_m)$.
In terms of $sl(2,\mathbb{R})\oplus u(1)$ generators, the basis transformation above corresponds to
\begin{eqnarray}\label{eq:IW-contraction-chiral}
L_{-1} =\alpha\mathcal{L}_{-1}\,,\quad
L_0 = \frac{1}{2}\mathcal{L}_0+\frac{\alpha^2}{2}\mathcal{N}_0\,, \quad
L_1 =\alpha\mathcal{N}_1\,, \quad
N_0 = -\frac{1}{2}\mathcal{L}_0+\frac{\alpha^2}{2}\mathcal{N}_0\,.
\end{eqnarray}
Then it follows that the coefficients of the $P_2^c$ and $sl(2,\RR)\oplus u(1)$ bilinear forms are related by
\begin{equation}
\gamma_s = \frac{1}{2}\gamma_1+\alpha^2\gamma_2\,,\qquad \gamma_u = \frac{1}{2}\gamma_1-\alpha^2\gamma_2\,,
\end{equation}
and likewise for the barred sector.
This means that at finite $\alpha$, the bilinear form satisfies
\begin{equation}
\label{eq:P11-trace-finite-alpha}
\langle \LL_{-1}, \mN_1 \rangle
= - \gamma_2 - \frac{\gamma_1}{2\alpha^2}, \quad
\langle \LL_0, \mN_0 \rangle
= \gamma_2, \quad
\langle \LL_0, \LL_0 \rangle
= \frac{\gamma_1}{2}, \quad
\langle \mN_0, \mN_0 \rangle
= \frac{\gamma_1}{2\alpha^4}.
\end{equation}

In the above, we have discussed the limit of $so(2,2)\oplus u(1) \oplus u(1)$ as a conformal symmetry algebra.
There is also a bulk perspective on this contraction.
Define $\mathcal{T}_a$ and $\mathcal{R}_a$
by
\begin{equation}
  \label{eq:eNH-bulk-isometry-basis}
\mathcal{T}_a=\frac{1}{2}\left(\mathcal{P}_a+\mathcal{K}_a\right)\,,\qquad \mathcal{R}_a=\frac{1}{2}\left(\mathcal{P}_a-\mathcal{K}_a\right)\,.
\end{equation}
Rescaling $\mathcal{T}_a$, $\mathcal{D}$ and $\mathcal{N}$ by $l$, the algebra \eqref{eq:contractedalgebra} can be written as
\begin{eqnarray}\label{eq:bulk-contracted-algebra}
  &&[\mathcal{T}_a,\mathcal{R}_b]=i\mathcal{N}\eta_{ab}\,,\qquad [\mathcal{M},\mathcal{T}_a]=i\epsilon_a{}^b\mathcal{T}_b\,,\qquad [\mathcal{M},\mathcal{R}_a]=i\epsilon_a{}^b\mathcal{R}_b\,,\nonumber\\
  &&[\mathcal{D},\mathcal{R}_a]=i\mathcal{T}_a\,,\qquad [\mathcal{D},\mathcal{T}_a]=\frac{i}{l^2}\mathcal{R}_a\,,\label{eq:contractedalgebra2}\\
  &&[\mathcal{T}_a,\mathcal{T}_b]=-\frac{i}{l^2}\mathcal{S}\epsilon_{ab}\,,\qquad [\mathcal{R}_a,\mathcal{R}_b]=i\mathcal{S}\epsilon_{ab}\,.\nonumber
\end{eqnarray}
Since the generator $\mS$ corresponds to a central extension, and since $\eta_{ab}$ has Minkowski signature,
we will refer to this algebra as the extended pseudo-Newton--Hooke algebra.
For $l\rightarrow\infty$, the first two lines of this algebra provide a different real form of the complexified Bargmann algebra\footnote{
  The Bargmann algebra would be obtained by replacing the 2D Minkowski metric $\eta_{ab}$ with the Euclidean metric $\delta_{ab}$.
}.
The analogy with the Bargmann algebra becomes clear if we view $\mathcal{D}$ as the Hamiltonian, $\mathcal{M}$ as the generator of rotations, $\mathcal{T}_a$ as the momenta, $\mathcal{R}_a$ as the Galilei boosts and $\mathcal{N}$ as the mass generator.
For finite $l$, the first two lines of \eqref{eq:bulk-contracted-algebra} are a different real form of the Newton--Hooke algebra\footnote{
The contraction we take is not one of the kinematical algebras classified by Bacry and Leblond (see also \cite{Figueroa-OFarrill:2017sfs,Figueroa-OFarrill:2017ycu, Figueroa-OFarrill:2017tcy} for a recent classification).
  Instead, it is a different real form of the complexified extended Newton--Hooke algebra, which is isomorphic to two copies of $E_2^c$ as opposed to $P_2^c$.
  The Chern--Simons theory for extended Newton--Hooke has been studied in \cite{Hartong:2016yrf,Papageorgiou:2010ud}}.
The addition of the central element $\mathcal{S}$
ensures that we can construct a non-degenerate invariant bilinear form on the algebra.

\subsection{Chern--Simons action after contraction}
\label{ssec:contractCSaction}
As we review in Appendix \ref{app:so22-cs-review}, the Chern--Simons Lagrangian with gauge algebra $so(2,2)$ reproduces the three-dimensional Einstein-Hilbert Lagrangian with cosmological constant \cite{Witten:1988hc}.
We now want to add the $u(1)$ connections, whose components we parametrize by
\begin{equation}\label{eq:betaN}
\mathbf{A}_{u(1)}=Z^1Q_1+Z^2Q_2=U N_0+\bar U\bar N_0\,.
\end{equation}
The relation between $(Q_1,Q_2)$ and $(N_0,\bar N_0)$ is given in \eqref{eq:N0-bar-N0-definition}.
Using the bilinear form in \eqref{eq:metricsl2+U1} their contribution to the CS Lagrangian is
\begin{align}\label{eq:u(1)2action}
\left\langle \mathbf{A}_{u(1)}, d\mathbf{A}_{u(1)} \right\rangle
  = \frac{\gamma_u-\bar{\gamma}_u}{2}\left(\frac{1}{l^2}Z^1\wedge dZ^1+Z^2\wedge dZ^2\right)+\frac{\gamma_u+\bar{\gamma}_u}{l}Z^1\wedge dZ^2.
\end{align}
The total $so(2,2)\oplus u(1)\oplus u(1)$ connection then consists of the following components,
\begin{equation}\label{eq:uncontr-gauge-field}
\mathbf{A}
= E^A T_A + \Omega^A J_A + Z^1 Q_1 + Z^2 Q_2.
\end{equation}
Recall that $E^A$ is the vielbein, $\Omega^A$ is the spin connection and $Z_1$ and $Z_2$ are two bulk $u(1)$ gauge fields.
In terms of these components, the total Chern--Simons Lagrangian is
\begin{align}
  \mathcal{L}_\text{CS}
    &= \frac{\gamma_s +\bar\gamma_s}{2l}
      \left(
        2 E^A \wedge d \Omega^B \eta_{AB}
        +\epsilon_{ABC}  E^A \wedge\Omega^B \wedge \Omega^C
        + \frac{1}{3l^2}
        \epsilon_{ABC} E^A \wedge E^B \wedge E^C
      \right)\nonumber\\
    \label{eq:app-general-so22-u12-cs-action}
    &{}\qquad
    + \frac{\gamma_s -\bar\gamma_s}{2}
      \left(
        \Omega^A \wedge d \Omega^B \eta_{AB}
        + \frac{1}{3} \epsilon_{ABC}\Omega^A \wedge \Omega^B \wedge \Omega^C+ \frac{1}{l^2}E^A \wedge d E^B \eta_{AB}\right.\\
  &{}\qquad\left.
        + \frac{1}{l^2} \epsilon_{ABC}
        E^A \wedge E^B \wedge \Omega^C
      \right)
    + \frac{\gamma_u-\bar{\gamma}_u}{2}\left(\frac{1}{l^2}Z^1\wedge dZ^1+Z^2\wedge dZ^2\right)
    + \frac{\gamma_u+\bar{\gamma}_u}{l}Z^1\wedge dZ^2.
    \nonumber
\end{align}

Now let us determine the relation between the connection components in the standard algebra and the contracted algebra.
We use the $l$-rescaled bulk algebra in \eqref{eq:bulk-contracted-algebra} and also scale the $Q_1$ generator correspondingly.
The contracted connection can then be written as
\begin{equation}\label{eq:contr-gauge-field}
\mathbf{A}
= \tau\mathcal{D} + e^a \mathcal{T}_a
+ m \mathcal{N} + \omega \mathcal{M}
+ \omega^a \mathcal{R}_a + \zeta \mathcal{S}.
\end{equation}
Here, $\tau$ and $e^a$ are Newton--Cartan vielbeine and $\omega^a$ and $\omega$ play the role of boost/spin connections.
The components $m$ and $\zeta$ correspond to the central extensions of the algebra.
Taking into account that $\mT_a$, $\mD$ and $\mN$ have been rescaled by a factor of $l$, the basis transformation in \eqref{eq:genredef} then leads to the following identifications,
\begin{equation}
\label{eq:vielbein-redefinitions}
\begin{gathered}
E^2 =\tau + \frac{m}{2\alpha^2}\,,\quad
E^a = \frac{1}{\alpha}\epsilon_b{}^a e^b\,,\quad
Z^1 = -\tau + \frac{m}{2\alpha^2}\,,\\
\Omega^2 = \omega + \frac{\zeta}{2\alpha^2}\,,\quad
\Omega^a = \frac{1}{\alpha}\omega^a\,,\quad
Z^2 = -\omega +\frac{\zeta}{2\alpha^2}\,.
\end{gathered}
\end{equation}
Using this in the action \eqref{eq:app-general-so22-u12-cs-action} and taking the limit $\alpha\to \infty$, the action on the contracted algebra $P_2^c\times P_2^c$ becomes
\begin{eqnarray}
\mathcal{L} & = & \frac{1}{2l}(\gamma_2+\bar\gamma_2)\left(2\tau\wedge d\zeta+2\epsilon_{ab}e^a\wedge d\omega^b+2m\wedge d\omega-\frac{1}{l^2}\tau\wedge\epsilon_{ab}e^a\wedge e^b\right.\nonumber\\
&&\left.+\tau\wedge\epsilon_{ab}\omega^a\wedge\omega^b+2\eta_{ab}e^a\wedge\omega^b\wedge\omega\right)\\
&&+\frac{1}{2}(\gamma_2-\bar\gamma_2)\left(\eta_{ab}\omega^a\wedge d\omega^b-\frac{1}{l^2}\eta_{ab}e^a\wedge de^b+\frac{2}{l^2}m\wedge d\tau+2\omega\wedge d\zeta\right.\nonumber\\
&&\left.+\frac{2}{l^2}\eta_{ab}\tau\wedge e^a\wedge\omega^b-\frac{1}{l^2}\epsilon_{ab}e^a\wedge e^b\wedge\omega+\omega\wedge\epsilon_{ab}\omega^a\wedge\omega^b\right)\nonumber\\
&&+\frac{1}{l}(\gamma_1+\bar\gamma_1)\tau\wedge d\omega+\frac{1}{2}(\gamma_1-\bar\gamma_1)\left(\frac{1}{l^2}\tau\wedge d\tau+\omega\wedge d\omega\right)\,.\nonumber
\end{eqnarray}
This is the CS action for two copies of $P_{2}^c$, using the connection \eqref{eq:contr-gauge-field}
and the metric \eqref{eq:p2cmetric}.
This action was first derived in \cite{Hartong:2016yrf} for extended Newton--Hooke, which is a different real form of the complexification of our contracted algebra.
Since the vielbein $e^a$ now involves a Lorentzian structure, we will refer to the bulk geometry as pseudo-Newton--Cartan gravity.

In the next section we will see that, in order for the limit to be properly defined on the full phase space of AdS$_3$ gravity, the distinguished vielbein $\tau$ has to correspond to the \emph{radial} direction.
In contrast to the usual Fefferman--Graham procedure in AdS, which is simply a choice of coordinates, the vielbein $\tau$ therefore defines an (absolute) radial foliation which is \emph{intrinsic} to the geometry.
It would be very interesting to investigate the consequences of this phenomenon on the RG flow of the corresponding field theories.

\subsection{Generalization to higher-dimensional algebra}
\label{ssec:higherdim}
The In\"on\"u--Wigner contraction of the 2D conformal algebra shown above can in fact be achieved for any dimension.
Consider the conformal algebra $so(d+1,2)$ in $d+1$ dimensions,
\begin{eqnarray} \nonumber
&& [D,P_a] =iP_a, \quad [D,K_a] = -iK_a, \quad [P_a,K_b] = -2iD \eta_{ab} -2iM_{ab}, \\ \nonumber
&& [M_{ab},K_c] =i\left(\eta_{ac} K_b- \eta_{bc} K_{a}\right), \quad [M_{ab},P_c] =i\left(\eta_{ac} P_b- \eta_{bc}P_a\right), \\
&& [M_{ab},M_{cd}] = i\left(\eta_{ac} M_{bd}+ \eta_{bd} M_{ac} - \eta_{ad} M_{bc} - \eta_{bc} M_{ad}\right)\,,
\end{eqnarray}
where $a,b=0,...,d$ and $\eta_{ab}$ is the Minkowski metric. We add to this a $u(1)$ generator $Q$. We can also introduce another $so(d,1)$ algebra generated by $Z_{ab}$ whose commutation relations are
\begin{equation}
[Z_{ab},Z_{cd}] =i\left(\eta_{ac} Z_{bd}+ \eta_{bd} Z_{ac} - \eta_{ad} Z_{bc} - \eta_{bc} Z_{ad}\right)\,.
\end{equation}
The total algebra is thus $so(d+1,2)\oplus u(1)\oplus so(d,1)$. In two spacetime dimensions we can write $Z_{ab}=Z\epsilon_{ab}$ and then $Q$ and $Z$ are the $Q_1$ and $Q_2$ generators of section \ref{sec:CSalgebra}.
Now let us make the following $\alpha$-dependent basis transformation,
\begin{eqnarray} \nonumber
&& P_a =\alpha \mathcal{P}_a, \quad K_a =\alpha \mathcal{K}_a, \quad D= \frac{\mathcal{D}}{2}+\alpha^2\mathcal{N}, \quad Q = \alpha^2\mathcal{N}-\frac{\mathcal{D}}{2}, \\
&& M_{ab} = \frac{\mathcal{M}_{ab}}{2} + \alpha^2 \mathcal{S}_{ab}, \quad Z_{ab} = \frac{\mathcal{M}_{ab}}{2} -\alpha^2 \mathcal{S}_{ab}\,.
\end{eqnarray}
We then see that for $\alpha\rightarrow\infty$ we obtain the algebra
\begin{eqnarray}\nonumber
&& [\mathcal{P}_a,\mathcal{K}_b] =-2i\mathcal{N} \eta_{ab} -2i\mathcal{S}_{ab}, \quad [\mathcal{D}, \mathcal{P}_a] = i\mathcal{P}_a, \quad [\mathcal{D},\mathcal{K}_a] =-i\mathcal{K}_a, \\ \nonumber
&& [\mathcal{M}_{ab},\mathcal{K}_c] =i\left(\eta_{ac} \mathcal{K}_b- \eta_{bc} \mathcal{K}_{a}\right), \quad [\mathcal{M}_{ab},\mathcal{P}_c] =i\left(\eta_{ac} \mathcal{P}_b- \eta_{bc}\mathcal{P}_{a}\right),  \\
&& [\mathcal{M}_{ab},\mathcal{M}_{cd}] =i\left( \eta_{ac} \mathcal{M}_{bd}+ \eta_{bd} \mathcal{M}_{ac} - \eta_{ad} \mathcal{M}_{bc} - \eta_{bc} \mathcal{M}_{ad} \right),\\
&& [\mathcal{M}_{ab},\mathcal{S}_{cd}] =i\left(\eta_{ac} \mathcal{S}_{bd}+ \eta_{bd} \mathcal{S}_{ac} - \eta_{ad} \mathcal{S}_{bc} - \eta_{bc} \mathcal{S}_{ad}\right)\,.\nonumber
\end{eqnarray}
If we had not included the $Z_{ab}$ generators we would have found the algebra that is obtained by setting $\mathcal{S}_{ab}=0$.
The latter algebra is the scaling nonconformal algebra in general dimensions.

Importantly, the contraction of $so(d+1,2)\oplus u(1)$ leads to a $(d+1)$-dimensional relativistic algebra with dilatation generators but no conformal generators. Instead of the conformal generators we have the $\mathcal{K}_a$ generators and a central element $\mathcal{N}$.
By defining
\begin{equation}
\mathcal{T}_a=\frac{1}{2}\left(\mathcal{P}_a +\mathcal{K}_a\right),\quad \mathcal{R}_a=\frac{1}{2}\left(\mathcal{P}_a -\mathcal{K}_a\right)
\end{equation}
and by rescaling $\mathcal{T}_a$, $\mathcal{N}$ and $\mathcal{D}$ using a length scale $l$,
we obtain the algebra
\begin{eqnarray}\nonumber
  \label{eq:higher-dim-algebra}
&& [\mathcal{T}_a,\mathcal{R}_b] =i\mathcal{N} \eta_{ab}, \quad [\mathcal{D},\mathcal{T}_a] =\frac{i}{l^2}\mathcal{R}_a, \quad [\mathcal{D},\mathcal{R}_a] =i\mathcal{T}_a, \\
&& [\mathcal{M}_{ab},\mathcal{T}_c] =i\left(\eta_{ac} \mathcal{T}_b- \eta_{bc} \mathcal{T}_{a}\right), \quad [\mathcal{M}_{ab},\mathcal{R}_c] =i\left(\eta_{ac} \mathcal{R}_b- \eta_{bc}\mathcal{R}_{a}\right)\,,  \\  \nonumber
&& [\mathcal{M}_{ab},\mathcal{M}_{cd}] =i\left( \eta_{ac} \mathcal{M}_{bd}+ \eta_{bd} \mathcal{M}_{ac} - \eta_{ad} \mathcal{M}_{bc} - \eta_{bc} \mathcal{M}_{ad} \right)\,,
\end{eqnarray}
For $l\rightarrow\infty$ this is a different real form of the complexified Bargmann algebra in $d$ dimensions.

The algebra \eqref{eq:higher-dim-algebra} is the $d$-dimensional generalization of \eqref{eq:bulk-contracted-algebra} without the central element $\mS$.
Gauging this algebra leads to $d$-dimensional pseudo-Newton--Cartan geometry.
It is tempting to speculate that \eqref{eq:higher-dim-algebra} governs the bulk gravity theory for near-BPS limits of CFTs in any dimension.

\section{Phase space of the limit theory}
\label{sec:phasespace}
Asymptotically AdS$_3$ spacetimes can be described using the $sl(2,\RR)$ Wess-Zumino-Witten (WZW) theory corresponding to $sl(2,\RR)$ Chern--Simons theory, see Appendix \ref{app:so22-cs-review} or \cite{Donnay:2016iyk} for a review.
In this section, we study the limit of the phase space corresponding to the WZW model of $so(2,2)\oplus u(1) \oplus u(1)$ and show that it reproduces the full $P_{2}^c\oplus P_{2}^c$ WZW phase space.
We then write down the metric components in terms of the original WZW currents.
Next, we study the symmetries of the limit of the Poincar\'e and global AdS$_3$ vacua and show that these vacua can be written as a coset space.
Finally, we show that the full contracted symmetry algebra, including the central extensions, can be realized on a scalar field coupled to a pseudo-Newton--Cartan background.

\subsection{Mapping relativistic phase space to contracted phase space}\label{ssection:metricdatap2c}
In appendix \ref{app:so22-cs-review}, we review the standard parametrization of the classical phase space of $so(2,2)$ Chern--Simons theory on a manifold with a boundary in terms of the currents of a WZW model.
This involves a choice of radial component $\mathbf{A}_\rho$ as well as a choice of chirality and results in the expression \eqref{eq:app-general-so22-connection-currents}, which reads
\begin{equation}
	\mathbf{A} = e^{i\rho(L_0 +\bar L_0)} \left(
				id + F^a(\xp) L_a d\xp + \bar{F}^a(\xm) \bar L_a\, d\xm
			\right)e^{-i\rho(L_0 +\bar L_0)}.
\end{equation}
The WZW phase space is parametrized by the three current components $F^a(\xp)$.
We now want to include two $u(1)$ factors.
Following Appendix \ref{app:variational-problem}, we also have to choose a chirality for these generators.
In terms of the generators $(Q_1,Q_2)$ and $(N_0,\bar{N}_0)$ introduced in \eqref{eq:betaN}, we choose the following parametrization of the $u(1)\oplus u(1)$ connection,
\begin{gather}
	U = F^Nd\xp+U_\rho d\rho\,,\qquad\bar
  U =\bar F^{\bar N}d\xm +\bar U_\rho d\rho, \\
  U N_0 + \bar U \bar N_0
    = Z^1 Q_1 + Z^2 Q_2.
\end{gather}
Here, we have defined
\begin{gather}
  Z^1_\rho
    = \frac{l}{2}\left(U_\rho+\bar U_\rho\right)\,,\qquad
  Z^2_\rho
    = \frac{1}{2}\left(U_\rho-\bar U_\rho\right)\,,\\
  Z^1 = \frac{l}{2}\left(F^N d\xp + \bar F^{\bar N}d\xm\right)
          + Z^1_\rho d\rho, \quad
  Z^2 = \frac{1}{2}\left(F^N d\xp - \bar F^{\bar N}d\xm\right)
          + Z^2_\rho d\rho.
\end{gather}
We allow for a contribution $U_\rho N_0 + \bar U_\rho \bar N_0$ coming from $u(1)\oplus u(1)$ in the radial component of the connection.
We will see below that such a contribution is crucial in order to have a well-defined limit of the phase space.

We would like the full WZW phase space to be finite and nonzero in the $\alpha\to\infty$ limit.
This can be achieved by scaling the currents $F^a(\xp)$ and $\bar F^a(\xp)$ appropriately.
For clarity, we focus on the unbarred sector in the following,
\begin{equation}
  A = e^{i\rho L_0}\left(
      id + \left(
        F^+ L_+ + F^0 L_0 + F^- L_-
        \right) d\xp
    \right) e^{-i\rho L_0}
    + F^N N_0 d\xp + U_\rho N_0 d\rho.
\end{equation}
Plugging in the algebra redefinition \eqref{eq:IW-contraction-chiral} for $sl(2,\RR)\oplus u(1)$,
\begin{equation*}
  L_- = \alpha \LL_-, \quad
  L_+ = \alpha \mN_+, \quad
  L_0 = \frac{1}{2} \LL_0 + \frac{\alpha^2}{2} \mN_0, \quad
  N_0 = -\frac{1}{2} \LL_0 + \frac{\alpha^2}{2} \mN_0,
\end{equation*}
the resulting connection can be written as
\begin{equation}
\begin{split}
  A &=
      \frac{1}{2} \left(1 - U_\rho\right) \LL_0 d\rho
      + \frac{\alpha^2}{2} \left(1 + U_\rho\right) \mN_0 d\rho\\
    &{}\qquad
      + \left(
        e^\rho \alpha F^+ \mN_+
        + \frac{1}{2}\left(F^0 - F^N\right) \LL_0
        + \frac{\alpha^2}{2} \left(F^0 + F^N\right) \mN_0
        + e^{-\rho} \alpha F^- \LL_-
      \right) d\xp.
\end{split}
\end{equation}
For $A$ to be finite as $\alpha\to\infty$, we have to set $U_\rho=-1$.
This means that the radial component of this $sl(2,\RR)\oplus u(1)$ factor of the full connection has to be set to
\begin{equation}
  A_\rho
    = \LL_0
    = L_0 - N_0.
\end{equation}
Furthermore, the following combinations have to be finite in the $\alpha\to\infty$ limit,
\begin{equation}
  \label{eq:curly-currents-definition}
  \mF^+ := \alpha F^+, \quad
  \mF^0 := \frac{1}{2} \left(F^0 - F^N\right), \quad
  \mF^- := \alpha F^-, \quad
  \mF^N := \frac{\alpha^2}{2} \left(F^0 + F^N\right).
\end{equation}
With those redefinitions, the connection is now
\begin{align}
	A
		&= \LL_0 d\rho
			+ e^\rho \mF^+ \mN_+ d\xp
			+ \mF^0 \LL_0 d\xp + \mathcal{F}^N \mN_0 d\xp
			+ e^{-\rho} \mF^- \LL_- d\xp\nonumber\\
		&= e^{i\rho \LL_0}\left(
				id + \left(
					\mF^+ \mN_+ + \mF^0 \LL_0 + \mathcal{F}^N \mN_0 + \mF^- \LL_-
					\right) d\xp
			\right) e^{-i\rho \LL_0}.\label{eq:P2connection}
\end{align}
This is the most general $P_{2}^c$ connection, which we obtain as a limit of the most general $sl(2,\RR)\oplus u(1)$ connection!
In other words, we can reach the full phase space of the $P_2^c$ theory from a limit of the phase space of the $sl(2,\RR)\oplus u(1)$ theory.
The total radial component is
\begin{equation}
  \label{eq:total-radial-component-choice}
  \mathbf{A}_\rho
    = \LL_0 + \bar\LL_0
    = L_0 + \bar L_0 - N_0 - \bar N_0
    = L_0 + \bar L_0 - l Q_1,
  \qquad
  Z^1_\rho = -l, \quad Z^2_\rho = 0.
\end{equation}
This background radial chemical potential offsets the dilatation generator $L_0+\bar L_0$ and leads to the new dilatation generator $\LL_0 + \bar\LL_0$.

Let us now study what the limit of the $so(2,2) \oplus u(1) \oplus u(1)$ Chern--Simons  phase space looks like in terms of the metric data.
In Appendix \ref{app:so22-cs-review} we recall how metric data maps to $so(2,2)$ connections, which are flat if the metric is on-shell.
The map from the most general $sl(2,\RR)\oplus sl(2,\RR)$ current components $F^a$ introduced in \eqref{eq:app-general-so22-connection-currents} to the vielbein and spin connection data is laid out in \eqref{eq:app-general-so22-vielbein-spin-connection-data}.

We now want to show explicitly that the identification of the connection components in \eqref{eq:vielbein-redefinitions} implies that the components of the contracted connection are nonzero and finite in the $\alpha\to\infty$ limit.
The vielbeine $e^a$ on the leaves of the foliation are given by
\begin{subequations}
    \label{eq:currents-to-eNH-data-vielbein}
\begin{align}
	{e}^0
		&= \alpha E^1
		= \frac{\alpha l}{2}
		\left(
			- e^\rho \left( F^+ d\xp - \FBp d\xm \right)
			+ e^{-\rho} \left( F^- d\xp - \FBm d\xm \right)
		\right) \nonumber \\
    &= \frac{l}{2}
    \left(
      - e^\rho \left( \mF^+ d\xp - \mFBp d\xm \right)
      + e^{-\rho} \left( \mF^- d\xp - \mFBm d\xm \right)
    \right),\\
	{e}^1
		&= \alpha E^0
		= \frac{\alpha l}{2}
		\left(
			e^\rho \left( F^+ d\xp + \FBp d\xm \right)
			+ e^{-\rho} \left( F^- d\xp + \FBm d\xm \right)
		\right) \nonumber \\
  &= \frac{l}{2}
    \left(
      e^\rho \left( \mF^+ d\xp + \mFBp d\xm \right)
      + e^{-\rho} \left( \mF^- d\xp + \mFBm d\xm \right)
    \right).
\end{align}
\end{subequations}
Next, ${\tau}$ and ${m}$ can be expressed in terms of the currents as follows.
\begin{align}\nonumber
	\tau
		&= \frac{1}{2}\left( E^2 - Z^1 \right)
    = \frac{1}{2}\left[
				l d\rho
				+ \frac{l}{2} \left(
						F^0 d\xp + \bar{F}^0 d\xm
					\right)
				- Z^1
			\right] = ld\rho
				+ \frac{l}{2}\mF^0 d\xp
				+ \frac{l}{2}\mFB^0  d\xm, \\
    \label{eq:currents-to-eNH-data-m}
	{m}
		&= \alpha^2 \left( E^2 + Z^1 \right)
    = \alpha^2 \left[
			l d\rho
			+	\frac{l}{2} \left(
					F^0 d\xp + \bar{F}^0 d\xm
				\right)
				+ Z^1
			\right] = l \mathcal{F}^N d\xp
    + l \bar{\mathcal{F}}^{\bar N} d\xm.
\end{align}
Note that knowledge of the components we have listed so far specifies all functions $(F^a,\bar F^a)$.
In particular, the remaining components can be integrated out.
Explicitly, $\omega$ and $\zeta$ are
\begin{align}\nonumber
	{\omega}
		&= \frac{1}{2}\left( \Omega^2 - Z^2 \right)
    = \frac{1}{2}\left[
				\frac{1}{2}\left(
					F^0 d\xp - \bar{F}^0 d\xm
				\right)
        - Z^2
			\right] = \frac{1}{2} \mF^0 d\xp
    - \frac{1}{2} \mFB^0 d\xm, \\
    \label{eq:currents-to-eNH-data-zeta}
	{\zeta}
		&= \alpha^2 \left( \Omega^2 + Z^2 \right)
    = \alpha^2\left[
				\frac{1}{2}\left(
					F^0 d\xp - \bar{F}^0 d\xm
				\right)
        + Z^2
			\right]
  	= \mathcal{F}^N d\xp
      - \bar{\mathcal{F}}^{\bar N} d\xm.
\end{align}
Finally, the $\omega^a$ can be expressed as
\begin{subequations}
  \label{eq:currents-to-eNH-data-spacetime-omega}
\begin{align}
	{\omega}^0
		&= \alpha \Omega^0
    = \frac{1}{2}
			\left(
				e^\rho \left( \mF^+ d\xp - \mFBp d\xm \right)
				+ e^{-\rho} \left( \mF^- d\xp - \mFBm d\xm \right)
			\right),\\
	{\omega}^1
    &= \alpha \Omega^1
    = \frac{1}{2}
      \left(
        - e^\rho \left( \mF^+ d\xp + \mFBp d\xm \right)
        + e^{-\rho} \left( \mF^- d\xp + \mFBm d\xm \right)
        \right).
\end{align}
\end{subequations}
We thus see that all the metric data is manifestly finite in the $\alpha\to\infty$ limit.

\subsection{Solutions and Killing symmetries} \label{ssec:vacuumsymmetry}

We now want to study the limit of the Poincar\'e and global AdS$_3$ connections in more detail.
In particular, we want to find the equivalent of the Killing symmetries for the pseudo-Newton--Cartan metric data.

Let us use $M$ to denote a bulk spacetime world index.
Consider a local gauge transformation with parameter $\Lambda=\xi^M A_M+\Sigma$.
Since the translation generators couple to the vielbeine, $A$ contains all translation generators.
Without loss of generality, we can therefore choose $\xi^M$ such that $\Sigma$ contains \emph{no} translation generators.
For our contracted bulk algebra \eqref{eq:bulk-contracted-algebra} this means that $\Sigma$ can be parametrized by
\begin{equation}
\Sigma=\lambda^a \mathcal{R}_a+\lambda \mathcal{M}+\sigma \mN+\beta \mathcal{S} \,.
\end{equation}
On-shell, the connection $A$ then transforms as follows under $\Lambda$,
\begin{equation}
\delta A_M=\mathcal{L}_\xi A_M+\partial_M\Sigma-i[A_M,\Sigma]\,.
\end{equation}
We see that $\xi^M$ corresponds to a diffeomorphism of the three-dimensional base manifold, while $\Sigma$ generates internal transformations in the tangent bundle.
Using the connection components defined in \eqref{eq:contr-gauge-field}, we see that the Newton--Cartan metric data $\tau_M$, $m_M$ and $h_{MN}=-\eta_{ab}e^a_Me^b_N$  transforms as\footnote{
  Note that we use a nonstandard sign in the definition of $h_{MN}$ to compensate for the exchange of indices that occurs in \eqref{eq:currents-to-eNH-data-vielbein}.
  This exchange is ultimately due to the fact that we have defined $\mT_a=(P_a+K_a)/2l$ without the $\tensor{\epsilon}{_a^b}$ factor in the uncontracted generator $T_a=\tensor{\epsilon}{_a^b}(P_b+K_b)/2l$.
}
\begin{eqnarray}
\delta\tau_M & = &
  \mathcal{L}_\xi\tau_M\,,\label{eq:delta1}\\
\delta h_{MN} & = &
  \mathcal{L}_\xi h_{MN}-\lambda_a\left(e^a_M\tau_N+e^a_N\tau_M\right)\,,\\
\delta m_M & = &
  \mathcal{L}_\xi m_M+\partial_M\sigma+\lambda_a e^a_M\,. \label{eq:delta3}
\end{eqnarray}

First, consider the connection corresponding to pure Poincar\'e AdS$_3$, which has vanishing Virasoro and affine $u(1)$ currents.
The corresponding Newton--Cartan metric data is
\begin{equation}\label{eq:Poincare}
  \begin{split}
    &\tau  =\frac{dr}{r}\,,\quad m=0\,, \quad
      e^0 = \frac{dx}{r}\,, \quad e^1 = \frac{dt}{r},\\
    &h_{MN}dx^M dx^N
      = \left( e^0_M e^0_N - e^1_M e^1_N \right)dx^M dx^N
      = \frac{1}{r^2}\left(-dt^2+dx^2\right)\,.
  \end{split}
\end{equation}
The Killing vectors of this solution can be found by solving \eqref{eq:delta1}--\eqref{eq:delta3},
\begin{eqnarray}
\xi^t & = & ct+\mu x+ a^t + \frac{1}{2} b^t r^2\,,\quad
  \xi^x = cx+\mu t + a^x+\frac{1}{2}b^x r^2\,,\quad
  \xi^r = cr\,,\\
\lambda^0 & = & b^x r\,,\quad
  \lambda^1= - b^t r\,,\quad
  \sigma= b^t t - b^x x\,.
\end{eqnarray}
Here, $a^a$ and $b^a$ are arbitrary (constant) two-vectors.
Note that the internal transformations parametrized by $\lambda$ drop out.
They are internal Lorentz transformations of the frame defined by $e^a$ which are always present but not necessary for our present considerations, so we will set $\lambda=0$ in the following.

We can collect six Killing vectors in the following diffeomorphism generators,
\begin{equation}\label{eq:Pgens}
\mP_a=-i\partial_a\,,\quad \mathcal{K}_a=-i r^2\partial_a\,,\quad \mD=-i(r\partial_r+x^a\partial_a)\,,\quad \mM =-i \epsilon_a{}^b x^a\partial_b\,.
\end{equation}
They form a representation of the generators in \eqref{eq:contractedalgebra} without central elements in terms of bulk diffeomorphisms.
Here, $r$ is a radial coordinate and $x^a=(t,x)$ parametrize the boundary.
Using $x^\pm=x\pm t$ the Killing vectors split into two commuting sets,
\begin{eqnarray}
  &&\LL_{-1} = -i\partial_+\,,
  \quad
  \LL_0 = -i\left(x^+\partial_++\frac{r}{2}\partial_r \right)\,,
  \quad
  \mN_{1} = -ir^2\partial_-\,,\label{eq:KVs}\\
  &&\bar\LL_{-1} = -i\partial_-\,,
      \quad
    \bar\LL_0 = -i\left(x^-\partial_-+\frac{r}{2}\partial_r \right)\,,
      \quad
    \bar\mN_{1} = -i r^2\partial_+\,,
\end{eqnarray}
These are three out of four generators of $P_2^c$.
Note that the central elements are not visible at the level of Killing vectors.
As we will show in the next subsection, they only act as internal symmetries on fields.

Second, the contraction of global AdS$_3$ with vanishing $u(1)$ currents leads to
\begin{equation}\label{eq:global}
\tau=d\rho\,,\quad h_{MN}dx^M dx^N=-\cosh^2\rho d\tau^2+\sinh^2\rho d\varphi^2\,,\quad m=0\,,
\end{equation}
where $\varphi$ is periodic with period $2\pi$. As in Einstein gravity, this solution can be related to the Newton--Cartan metric data of the Poincar\'e solution we studied above.
Consider \eqref{eq:delta1}--\eqref{eq:delta3} and leave out the diffeomorphisms.
These transformations can be exponentiated to the following finite transformations,
\begin{eqnarray}
  \tau'_M & = &
    \tau_M\,,\nonumber\\
  m'_M & = &
    m_M+\partial_M\sigma
      + \lambda_ae^a_M
      + \frac{1}{2}\lambda_a\lambda^a\tau_M\,,\label{eq:PTNCtrafos}\\
  h'_{MN} & = &
    h_{MN}
      - \lambda_a e^a_M\tau_N
      - \lambda_a e^a_N\tau_M
      - \lambda_a\lambda^a\tau_M\tau_N\,.  \nonumber
\end{eqnarray}
We first perform the coordinate transformation
\begin{equation}\label{eq:PtoG}
t=\frac{1}{2}(R^2+1)\tau\,,\qquad x=\frac{1}{2}(R^2-1)\varphi\,,\qquad r=R
\end{equation}
where $R=e^\rho$ so that $h_{MN}dx^Mdx^N$ in \eqref{eq:Poincare} matches $h_{MN}dx^Mdx^N$ in \eqref{eq:global} up to terms involving $dR$.
Then we find a $\lambda^a$ such that the transformed $h'_{MN}dx^Mdx^N$ is exactly \eqref{eq:global}, which can be achieved by taking
\begin{equation}
\lambda^0 = -R\vphi\,,\qquad\lambda^1 = -R\tau\,.
\end{equation}
Finally, to make sure that the transformed $m$ connection remains equal to zero, we set $\sigma$ to
\begin{equation}
\sigma=\frac{1}{4}(R^2+1)\tau^2-\frac{1}{4}(R^2-1)\varphi^2\,.
\end{equation}

Now that we have related the pseudo-Newton--Cartan contractions of global and Poincar\'e AdS$_3$, we can also identify their Killing vectors.
Two manifest Killing vectors in the global vacuum \eqref{eq:global} are $\partial_\tau$ and $\partial_\varphi$.
Using the coordinate transformation \eqref{eq:PtoG} we see that the global AdS$_3$ time and cylinder rotation generators correspond to
\begin{equation}
\partial_\tau=\frac{1}{2}(r^2+1)\partial_t=\frac{1}{2}(\mathcal{K}_0+ \mathcal{P}_0)=\mathcal{T}_0\,,\qquad\partial_\varphi=\frac{1}{2}(r^2-1)\partial_x=\frac{1}{2}(\mathcal{K}_1-\mathcal{P}_1)=-\mathcal{R}_1\,.
\end{equation}
We see that $\mathcal{T}_0$ and $\mathcal{R}_1$ are the contracted versions of the global AdS$_3$ generators $T_1$ and $J_1$ defined in \eqref{eq:AdSspacegens}, respectively.
Note that the AdS algebra \eqref{eq:AdSspacegens} has an inner automorphism, corresponding to a rotation in the 1--2 plane, which sends $(T_0,T_1,T_2)\mapsto(T_0,T_2,-T_1)$ and $(J_0,J_1,J_2)\mapsto(J_0,J_2,-J_1)$.
In global AdS$_3$, $T_1$ is the generator of global AdS time.
The inner automorphism means that in the AdS$_3$ geometry, a Killing vector for $T_1$ can equally be viewed as a generator for dilatations $T_2=D/l$.
Put differently, there are coordinate transformations, generated by the isometry corresponding to the inner automorphism, that map global AdS$_3$ back to global AdS$_3$ and map $\partial_\tau$ from $T_1$ to $T_2$. This automorphism, which is important for the state-operator map in AdS/CFT, is no longer present after we take the contraction.

\subsubsection*{Coset description}
The vacuum solutions \eqref{eq:Poincare} and \eqref{eq:global} are homogeneous spacetimes.
Just as one can think of AdS$_3$ as the coset $SO(2,2)/SO(2,1)$, we could expect to be able to write these solutions as a coset space.
Indeed, this can be done using the description of nonrelativistic geometries as coset spaces recently studied in \cite{Grosvenor:2017dfs}.

For a general Lie group $\mathcal{G}$ with subgroup $\mathcal{H}$, we denote its coset by $\mathfrak{M}= \mathcal{G}/\mathcal{H}$.
We split the Lie algebra $\mathfrak{g}$ in the subalgebra $\mathfrak{h}$ and its complement $\mathfrak{m}$.
Note that $\mathfrak{m}$ is generically not a Lie algebra, and $\mathfrak{M}$ is generically not a Lie group.
Choose a basis of $\mathfrak{g}$ that splits in elements of the subalgebra $\mathfrak{h}$, which we denote by $T_I$, and elements of the coset $\mathfrak{m}$, which we denote by $T_a$.
We will use $I,J,K\ldots$ for indices of $\mathfrak{h}$ and $a,b,c,\ldots$ for indices of $\mathfrak{m}$.

The coset space $\mathfrak{M}$ is a manifold of dimension $|\mathfrak{m}|$, which we can parametrize using coordinates $x^a$.
Now choose a coset representative
\begin{equation}
g = \prod_{a=0}^{|\mathfrak{m}|-1}\exp\left(x^a T_a\right) \in \mathcal{G}.
\end{equation}
To construct an $\mathcal{H}$-invariant metric on $\mathfrak{M}$, we have to find a symmetric bilinear form $\Omega$ on $\mathfrak{g}/\mathfrak{h}$ that is invariant under the adjoint action of $\mathfrak{h}$.
If we want to describe non-Riemannian geometries, the bilinear form $\Omega$ is degenerate and the corresponding construction has been worked out in \cite{Grosvenor:2017dfs}.
Instead of a non-degenerate bilinear form, one has to use a \emph{pair} of degenerate bilinear forms $(\Omega_{ab},\Omega^{ab})$ that are $\mathfrak{h}$-invariant,
\begin{equation}
f_{aI}^{\ \ c} \Omega_{cb} +f_{bI}^{\ \ c} \Omega_{ca} =0, \qquad \Omega^{ac} f_{cI}^{\ \ b} + \Omega^{bc} f_{cI}^{\ \ a} =0\,.
\end{equation}
It can be shown that for an appropriate choice of $\mathfrak{m}$ and $\mathfrak{h}$, the degenerate pair of bilinear forms $(\Omega_{ab},\Omega^{ab})$ on the coset $\mathfrak{m}$ exactly induces the degenerate (pseudo)-Newton--Cartan metric on the coset $\mathfrak{M}$.
The form $\Omega_{ab}$, which turns out to be of rank one, is used to define the one-form $\tau$, while $\Omega^{ab}$ turns out to be rank two for three-dimensional spacetime and defines the inverse spatial metric $h^{\mu\nu}$.
For more details, we refer the reader to \cite{Grosvenor:2017dfs}.

For the case we are interested in we start with the group $\mathcal{G} = \mathbf{P}_2^c \otimes \mathbf{P}_2^c$.
To obtain a three-dimensional pseudo-Newton--Cartan manifold we have to quotient out a subgroup with five generators.
A natural candidate is $\mathbf{P}_2^c \otimes U(1)$ where $\mathbf{P}_2^c$ is the diagonal subgroup of $\mathbf{P}_2^c \otimes \mathbf{P}_2^c$.

Using the basis in \eqref{eq:contractedalgebra2}, the diagonal subgroup plus the $U(1)$ of $\mathbf{P}_2^c \otimes \mathbf{P}_2^c$ is generated by $\mathfrak{h} =\{ \mR_1,\mR_0,\mM,\mathcal{S},\mN\}$ while the coset generators are $\mathfrak{m} = \{\mD,\mT_0,\mT_1\}$\footnote{
  Here, we have used the automorphism $\mathcal{T}_0\leftrightarrow\mathcal{R}_1$, $\mathcal{D}\leftrightarrow\mathcal{M}$ and $\mathcal{N}\leftrightarrow\mathcal{S}$.
}.
The corresponding $\mathfrak{h}$-invariant bilinear forms are $\Omega_{ab}\propto\text{diag}\{1,0,0\}$ and $\Omega^{ab} \propto \text{diag}\{0,-1,1\}$.
By expanding the Maurer--Cartan 1-form $g^{-1}dg$ as follows
\begin{equation}
g^{-1}dg = iT_a e^a + iT_Im^I
\end{equation}
we can read off the coset vielbeine $e^a$. The metrics are then $\tau_M\tau_N=\Omega_{ab}e^a_Me^b_N$ and $h^{MN}=\Omega^{ab}e^M_ae^N_b$.
We can generalize this construction to higher dimensions using the contracted algebra in section \ref{ssec:higherdim}.

Now let us turn to the two solutions considered above.
For the Poincar\'e AdS$_3$ limit, we choose the following coset representative
\begin{equation}\label{eq:group1}
g= e^{i(\mT_1+\mR_1)x} e^{i(\mT_0+\mR_0)t} e^{i\mD \rho}\,.
\end{equation}
The Maurer-Cartan 1-form is
\begin{equation}
g^{-1} dg = e^{\rho} (\mT_1+\mR_1) idx +e^{\rho} (\mT_0+\mR_0)i dt +i \mD d\rho\,.
\end{equation}
The corresponding vielbeine are
\begin{equation}
\tau = d\rho, \quad e^0 = e^{\rho} dt, \quad e^1 = e^{\rho} dx\,.
\end{equation}
By defining $r=e^{-\rho}$, we reproduce the pseudo-Newton--Cartan geometry in Poincar\'e coordinates \eqref{eq:Poincare}.
If we instead choose the coset representative to be
\begin{equation}\label{eq:group2}
g= e^{i\mR_1\varphi} e^{i\mT_0 \tau} e^{i\mD \rho}
\end{equation}
we reproduce the limit of the global AdS$_3$ solution (see Appendix \ref{app:global-ads3-sol})
\begin{equation}
g^{-1}dg=i\mathcal{T}_0\cosh\rho d\tau+i\mathcal{T}_1\sinh\rho d\varphi+i\mathcal{D}d\rho+i\mathcal{R}_0\sinh\rho d\tau+i\mathcal{R}_1\cosh\rho d\varphi\,.
\end{equation}
All these solutions locally have an algebra of Killing symmetries isomorphic to the centerless two-dimensional Poincar\'e algebra in \eqref{eq:KVs}.
In order to see the central extensions, it is crucial to add matter fields which we now show using the simple example of a bulk scalar field.

\subsection{Central extension on bulk scalar field}
\label{ssec:bulk-scalar}
Consider the following bulk action of a complex scalar field coupled to pseudo-NC geometry,
\begin{equation}\label{eq:probe}
S=\int d^3x e\left[iv^M\left(\psi D_M\psi^*-\psi^\star D_M\psi\right)-h^{MN}D_M\psi D_N\psi^\star\right]\,.
\end{equation}
The covariant derivative is $D_M\psi=(\partial_M-im_M)\psi$ and $e$ is the determinant of the matrix $(\tau_M, e^a_M)$.
The inverse metric $h^{MN}$ is defined as $h^{MN}=\eta^{ab}e^M_a e^N_b$ where the square matrix $(v^M, e^M_a)$ is the inverse of $(\tau_M, e^a_M)$, so we have $v^M\tau_M=1$, $v^M e_M^a=0$, $e^M_a\tau_M=0$ and $e^M_a e_M^b=\delta_a^b$.
Using these inverse relations, \eqref{eq:PTNCtrafos} implies that $v^M$ and $h^{MN}$ transform as
\begin{equation}
h'^{MN}=h^{MN}\,,\qquad v'^M=v^M-\lambda^a e_a^M\,.
\end{equation}
This transformation together with \eqref{eq:PTNCtrafos} leaves the action \eqref{eq:probe} invariant if $\psi$ transforms as
\begin{equation}
\psi=e^{-i\sigma}\psi'\,.
\end{equation}
Since the action \eqref{eq:probe} is invariant under the local pseudo-NC gauge symmetries it is guaranteed that it is invariant under the Killing symmetries of the background geometry.
What is interesting is that some of these will act in a nontrivial manner on $\psi$ giving rise to central extensions of the algebra of Killing symmetries.
For example, if we evaluate \eqref{eq:probe} on the background \eqref{eq:Poincare}, we obtain the action
\begin{equation}\label{eq:probe2}
S=\int d\rho dx^- dx^+\left[ i\left(\psi\partial_\rho\psi^*-\psi^\star\partial_\rho\psi\right)-2e^{2\rho}\left(\partial_+\psi\partial_-\psi^\star+\partial_-\psi\partial_+\psi^\star\right)\right]\,.
\end{equation}
This action is invariant under the following transformations
\begin{eqnarray}
&&x'^\pm=x^\pm+a^\pm\,,\nonumber\\
&&\rho'=\rho+\frac{1}{2}\log\lambda_+\,,\qquad x'^+=\lambda_+ x^+\,,\qquad\psi(\rho,x^-,x^+)=\lambda_+^{1/2}\psi'(\rho',x'^-,x'^+)\,,\nonumber\\
&&\rho'=\rho+\frac{1}{2}\log\lambda_-\,,\qquad x'^-=\lambda_- x^-\,,\qquad\psi(\rho,x^-,x^+)=\lambda_-^{1/2}\psi'(\rho',x'^-,x'^+)\,,\label{eq:symmetriesscalar}\\
&&x'^+=x^++e^{2\rho}v^+\,,\qquad\psi(\rho,x^-,x^+)=e^{iv^+x'^-}\psi'(\rho',x'^-,x'^+)\,,\nonumber\\
&&x'^-=x^-+e^{2\rho}v^-\,,\qquad\psi(\rho,x^-,x^+)=e^{iv^-x'^+}\psi'(\rho',x'^-,x'^+)\,,\nonumber\\
&&\psi(\rho,x^-,x^+)=e^{iq}\psi'(\rho,x^-,x^+)\,,\nonumber
\end{eqnarray}
where $a^\pm$, $\lambda_\pm$, $v^\pm$ and $q$ are constants. These transformations form the group $\mathbf{P}_2^c\otimes \mathbf{P}_2^c$ where the two extensions are identified. In other words we have under $\mathcal{N}$ and $\mathcal{S}$
\begin{equation}
\mathcal{N}\psi=\psi\,,\qquad \mathcal{S}\psi=0\,.
\end{equation}
The infinitesimal version of \eqref{eq:symmetriesscalar} is given by the Killing vectors \eqref{eq:KVs} together with an additional internal transformation of the form
\begin{equation}
\begin{split}
  &\mathcal{L}_{-1}\psi=-i\partial_+\psi\,,
  \quad
  \mathcal{L}_0\psi=-i\left(x^+\partial_++\frac{1}{2}\partial_\rho +\frac{1}{2}\right)\psi\,,
  \quad
  \mathcal{N}_1\psi=-i\left(e^{2\rho}\partial_-+ix^+\right)\psi\,,\\
  &\bar{\mathcal{L}}_{-1}\psi=-i\partial_-\psi\,,
    \quad
  \bar{\mathcal{L}}_{0}\psi=-i\left(x^-\partial_-+\frac{1}{2}\partial_\rho +\frac{1}{2}\right)\psi\,,
    \quad
  \bar{\mathcal{N}}_1\psi=-i\left(e^{2\rho}\partial_++ix^-\right)\psi\,.
\end{split}
\end{equation}

It would be interesting to see if this scalar field couples to an operator on the boundary and to work out the representation of this operator under the global symmetry group.

\section{Asymptotic symmetry algebras}
\label{sec:asymptalgebra}
In $so(2,2)$ Chern--Simons theory, the asymptotic Virasoro symmetries of AdS$_3$ can be understood in terms of a Drinfeld-Sokolov reduction of the corresponding boundary $sl(2,\RR)$ WZW model.
The reduction arises from a Dirichlet constraint on the leading radial components of the metric.
See Appendix \ref{app:cs-asymptotics-sl2-u1-decoupled} for a review.
In terms of the Newton--Hooke metric data, we can see from \eqref{eq:currents-to-eNH-data-vielbein} that constraining the leading-order behavior of $e^a$ corresponds to setting $\mF^+\equiv 1$.
The $P_2^c$ connection is then
\begin{equation}
		\label{eq:P11c-constrained-connection}
	a_+ = \mN_+ + \mF^0 \mL_0 + \mF^- \mL_- + \mF^N \mN_0.
\end{equation}
This restriction corresponds to adding a constraint $\chi$ to the initial Hamiltonian $H_0$,
\begin{equation}
	\chi = \int dx \, \Lambda(x) \left(\mF^+(x) - 1\right),
    \quad
	H_T = H_0 + \chi.
\end{equation}
Here, $\Lambda$ is an arbitrary function which acts as a Lagrange multiplier imposing $\mF^+(x)\equiv 1$.
This constraint generates gauge transformations through the Poisson bracket  \eqref{eq:app-current-poisson-bracket} on the WZW phase space,
\begin{equation}\label{eq:current-poisson-bracket}
  \{\mF^a(\xp), \mF^b(y^+)\}
    = - \frac{1}{2} \tensor{f}{^{ab}_c} \mF^c(\xp) \delta (\xp-y^+)
    + \frac{1}{2} \partial_{\xp} \delta (\xp-y^+) \Omega^{ab}.
\end{equation}
The structure constants are defined via $[T_a, T_b] = i\tensor{f}{_{ab}^c} T_c$ and indices should be raised and lowered using the invariant bilinear form $\Omega_{ab}=\langle T_a,T_b\rangle$.
Since we consider a nondegenerate form, its inverse exists and is denoted by $\Omega^{ab}$.
Using the bilinear form in \eqref{eq:P11-trace-finite-alpha} and the structure constants from \eqref{eq:P11-finite-alpha}, we find that the constraint $\chi$ generates the following gauge transformations on the currents in \eqref{eq:P11c-constrained-connection},
\begin{subequations}
		\label{eq:P11c-constraint-transformations}
\begin{align}
	\delta_\Lambda \mF^0(x)
		&= \{ \chi, \mF^0(x) \}
    = \frac{\Lambda(x)\mF^+(x)}{\gamma_1 + 2 \alpha^2 \gamma_2}
    \equiv \frac{\Lambda(x)}{\gamma_1 + 2 \alpha^2 \gamma_2}, \\
	\delta_\Lambda \mF^-(x)
		&= \{ \chi, \mF^-(x) \}
		= \frac{\partial \Lambda(x) + \Lambda(x)\mF^0(x) + \Lambda(x)\mF^N(x)/\alpha^2}
        {2\gamma_2 + \gamma_1/\alpha^2},\\
        	\delta_\Lambda \mF^N(x)
		&= \{ \chi, \mF^N(x) \}
		= \frac{\Lambda(x)\mF^+}{2\gamma_2 + \gamma_1/\alpha^2}
    \equiv \frac{\Lambda(x)}{2\gamma_2 + \gamma_1/\alpha^2}.
\end{align}
\end{subequations}
One could use these transformations to set part of the currents to zero.
However, to make sure that no information is lost in the contraction, we will not do any gauge fixing.
Instead, we will work with what we refer to as \emph{physical} currents, which are combinations of the currents in \eqref{eq:P11c-constrained-connection} that are invariant under the gauge transformations generated by the constraint.
This method is reviewed in appendix \ref{app:cs-asymptotics-sl2-u1-decoupled} for the reduction to Virasoro.

In this section, we will first compute the asymptotic symmetry algebra (ASA) at infinite contraction parameter $\alpha$ in section \ref{ssec:infinitealphaASG}.
This gives a \emph{warped} Virasoro-affine $u(1)$ algebra.
We then work out explicitly the finite $\alpha$ ASA in section \ref{ssec:finitealphaASG}, where we also show that its limit is well-defined and reproduces the $\alpha\to\infty$ result.

However, at finite $\alpha$ the Chern--Simons algebra is equivalent to $sl(2,\RR)\oplus u(1)$, as we discussed in the previous section.
Its natural ASA is an \emph{uncoupled} Virasoro-affine $u(1)$ algebra, as we demonstrate in section \ref{ssec:sectionuncouple}.
We then show explicitly how the coupled algebra at finite $\alpha$ can be obtained from a redefinition of the uncoupled algebra.

\subsection{Warped Virasoro algebra in limit theory}
\label{ssec:infinitealphaASG}
Only certain combinations of the $\mF^a$ in \eqref{eq:P11c-constrained-connection} correspond to a physical, conserved current with nontrivial boundary charges.
Such `physical' currents should be invariant under the gauge transformations in \eqref{eq:P11c-constraint-transformations}.
In the limit $\alpha\to\infty$, the following combinations are invariant,
\begin{equation}\label{eq:invariantblocks}
	 \mF^- - \mF^0 \mF^N - \partial \mF^N,
    \qquad
  \mF^0\,.
\end{equation}
The physical currents should be made up out of such invariant building blocks.
Recall that the infinitesimal boundary charges for Chern--Simons theory are given by \eqref{eq:app-cs-boundary-charges},
\begin{equation}
  \delta Q_\lambda
  = -2 \oint_{\partial\Sigma} \langle \lambda, \delta_\lambda a \rangle.
\end{equation}
We contracted the Chern--Simons algebra, but the manifold on which the Chern--Simons connections are defined is unchanged.
In other words, the fibers are different but the base manifold is still a cylinder, so we parametrize the boundary cycle $\partial\Sigma=S^1$ using a periodic coordinate $\vphi$.
For a gauge parameter $\lambda$ to preserve the constrained form of the connection in \eqref{eq:P11c-constrained-connection}, we need
\begin{equation}\label{eq:infinite-alpha-lambda0}
	\lambda^0 = \mF^0 \lambda^+ - \partial \lambda^+.
\end{equation}
Then the infinitesimal charge is given by
\begin{align}
	\delta Q_\lambda
    &= - 2\oint d\vphi \langle\lambda, \delta a_+\rangle
		= \oint d\vphi \left(
      - \gamma_1 \lambda^0 \delta \mF^0
  			+ 2\gamma_2 \left( \lambda^+ \delta \mF^- - \lambda^0 \delta \mF^N
          - \lambda^N \delta \mF^0 \right)
      \right)\\
		&= \oint d\vphi \left(
        \lambda^+ \left( -\gamma_1\left[\mF^0 \delta \mF^0 + \partial \delta \mF^0\right]
  				+ 2\gamma_2\left[\delta\mF^- - \mF^0\delta \mF^N - \delta\partial \mF^N\right]
  			\right)
  				- 2\gamma_2 \lambda^N \delta \mF^0
        \right). \nonumber
\end{align}
But now we have a problem: the $\mF^0\delta \mF^N$-term is not a total variation, so the infinitesimal charge is not integrable as it stands.
To fix this, we can define a new parameter $\bar\lambda^N$ via
\begin{equation}
	\label{eq:infinite-alpha-parameter-redefinition}
	\lambda^N = \bar\lambda^N + \mF^N \lambda^+.
\end{equation}
Then the infinitesimal charge integrand is integrable and can be written as
\begin{align}\nonumber
	-2\langle\lambda, \delta a_+\rangle
		&= \lambda^+ \,\delta \left(
				- \gamma_1 \left[ \frac{1}{2}(\mF^0)^2 + \partial \mF^0 \right]
				+ 2\gamma_2 \left[ \mF^- - \mF^0 \mF^N - \partial \mF^N \right]
			\right)
			- \bar\lambda^N \, \delta \left( 2\gamma_2 \mF^0 \right) \\ 	\label{eq:infinite-alpha-charge-integrand}
		&= \lambda^+ \delta \mT + \bar\lambda^N \delta \mJ.
\end{align}
Here, we have defined the two physical currents $\mT$ and $\mJ$,
\begin{align}
	\mT
		&= - \gamma_1 \left[ \frac{1}{2}(\mF^0)^2 + \partial \mF^0 \right]
			+ 2\gamma_2 \left[ \mF^- - \mF^0 \mF^N - \partial \mF^N \right], \\
	\mJ
		&= - 2\gamma_2 \mF^0.
\end{align}
Indeed, these currents are composed out of the combinations in \eqref{eq:invariantblocks} and are therefore invariant under the constraint gauge transformations.
Under the residual transformations satisfying \eqref{eq:infinite-alpha-lambda0}, the physical currents transform as follows,
\begin{align}
  \delta \mT
    &= \lambda^+ \pd \mT + 2\mT \pd \lambda^+
      + \gamma_1 \pd^{3}\lambda^+
      + \mJ \pd \bar\lambda^N - 2\gamma_2 \pd^{2}\bar\lambda^N, \\
  \delta \mJ
    &= \lambda^+ \pd \mJ + \mJ \pd \lambda^+
      + 2\gamma_2 \pd^{2} \lambda^+.
\end{align}
The Poisson bracket of the boundary charges can be determined using the WZW Poisson bracket \eqref{eq:current-poisson-bracket}.
First, it is useful to split $Q_\lambda$ into a Virasoro and affine $u(1)$ charges,
\begin{equation}
Q_\lambda
  = \oint d\varphi\left(\lambda^+\mT + \bar\lambda^N\mJ\right)
  = Q^{\text{Vir}}[\lambda^+] + Q^{u(1)}[\bar\lambda^N]\,.
\end{equation}
They satisfy the following algebra,
\begin{align}
  \{Q^{\text{Vir}}[\lambda^+],Q^{\text{Vir}}[\mu^+]\}
    &= Q^{\text{Vir}}\left[\mu^+\partial\lambda^+-\lambda^+\partial\mu^+\right]
      + \gamma_1\oint d\varphi\mu^+\partial^3\lambda^+\,,\\
  \{Q^{\text{Vir}}[\lambda^+],Q^{u(1)}[\bar\mu^N]\}
    &= -Q^{u(1)}\left[\lambda^+\partial\bar\mu^N\right]
      + 2\gamma_2\oint d\varphi\bar\mu^N\partial^2\lambda^+\,.
\end{align}
Following \eqref{eq:app-virasoro-modes}, if we expand the Virasoro and affine $u(1)$ charges in terms of the modes
\begin{equation}
  \mathcal{L}_m = - Q^{\text{Vir}}[e^{im\varphi}]\,,
    \qquad
  \mathcal{N}_m = - Q^{u(1)}[e^{im\varphi}]\,,
\end{equation}
we find that these modes satisfy the following commutation relations
\begin{align}
  \{\LL_m, \LL_n\}
    &= -i(m-n) \LL_{m+n} - 2\pi i \gamma_1 m^3 \delta_{m+n,0}, \\
  \{\LL_m, \mN_n\}
    &= in \mN_{m+n} - 4\pi \gamma_2 m^2 \delta_{m+n,0}.
\end{align}
Replacing $i\{\cdot,\cdot\}$ by $[\cdot,\cdot]$, and shifting the zero modes of the algebra using $\LL_m\to \LL_m + \pi \gamma_1\delta_{m,0}$ and $\mN_m\to\mN_m + 4\pi\gamma_2 i\delta_{m,0}$, this yields
\begin{subequations}
  \label{eq:P11-infinite-alpha-ASA}
\begin{align}
  [\LL_m, \LL_n]
    &= (m-n) \LL_{m+n} + 2\pi \gamma_1 m(m^2-1) \delta_{m+n,0}, \\
  [\LL_m, \mN_n]
    &= - n \mN_{m+n} - 4\pi i \gamma_2 m(m+1) \delta_{m+n,0}.
\end{align}
\end{subequations}
This is a warped Virasoro algebra with an extension in the Virasoro-affine $u(1)$ commutator and vanishing level of the affine $u(1)$.
It appeared before in the context of Rindler holography \cite{Afshar:2015wjm} with vanishing Virasoro central charge.
Here, we find it using a systematic Drinfeld-Sokolov reduction of a $P_2^c$ WZW model.

\subsection{Asymptotic symmetries from contraction}\label{ssec:finitealphaASG}
In fact we can do more.
As we explained for the Chern--Simons algebra in Section \ref{sec:CSalgebra}, the In\"on\"u-Wigner-type contraction is nothing but a basis transformation until we take $\alpha\to\infty$.
We now want to demonstrate that this is also true on the level of the asymptotic symmetry algebra.
To do that we first repeat the above computation at finite $\alpha$.
Again we start with
\begin{equation}\label{eq:asym-p2c-sol}
  a_+ = \mN_+ + \mF^0 \mL_0 + \mF^- \mL_- + \mF^N \mN_0.
\end{equation}
We now use the finite $\alpha$ commutation relations from \eqref{eq:P11-finite-alpha}.
The constrained connection is then preserved by
\begin{equation}
\label{eq:finite-alpha-allowed-transformation}
	\lambda^0
		= \left(\mF^0 - \frac{\mF^N}{\alpha^2}\right) \lambda^+
      - \partial \lambda^+
      - \frac{\lambda^N}{\alpha^2}.
\end{equation}
Now let us write out the expression for infinitesimal charges.
As before, we will see that a redefinition of $\lambda^N$ will be necessary to obtain an integrable expression.
Using the bilinear form at finite $\alpha$ in \eqref{eq:P11-trace-finite-alpha}, we find
\begin{align}
  \delta Q_\lambda
    &= -2\oint d\vphi\left\langle\lambda, \delta a_+\right\rangle\\
		&= -2 \oint d\vphi\left(
        - \lambda^+
          \left( \gamma_2 + \frac{\gamma_1}{2\alpha^2} \right) \delta \mF^-
  			+ \lambda^0
          \left( \gamma_2 \delta \mF^N + \frac{\gamma_1}{2} \delta \mF^0 \right)
  			+ \lambda^N \left( \frac{\gamma_1}{2\alpha^4} \delta \mF^N
  				+ \gamma_2 \delta \mF^0 \right)
        \right)
      \nonumber
      \\
		&= -2 \oint d\vphi \, \lambda^+ \left[
				- \left( \gamma_2 + \frac{\gamma_1}{2\alpha^2} \right) \delta \mF^-
				+ \left( \mF^0 + \frac{\mF^N}{\alpha^2} \right)
					\left( \gamma_2 \delta \mF^N + \frac{\gamma_1}{2} \delta \mF^0 \right)
				+ \gamma_2 \delta \partial \mF^N
				+ \frac{\gamma_1}{2} \delta \partial \mF^0
			\right] \nonumber \\
		&{}\qquad \nonumber
			-2 \oint d\vphi\, \lambda^N
        \left( \gamma_2 - \frac{\gamma_1}{2\alpha^2} \right)
				\left( \delta \mF^0 - \frac{\delta \mF^N}{\alpha^2} \right).
\end{align}
Again, we see that the infinitesimal charge is not integrable, which we fix by setting
\begin{equation}
	\lambda^+ = \lambda^+,
		\quad
	\lambda^N = \bar\lambda^N + \mF^N \lambda^+.
\end{equation}
The boundary charges can then be written as
$
  Q_\lambda
    = \oint d\vphi \left(\lambda^+ \mT + \bar\lambda^N \mJ\right)
$
with currents
\begin{align}
  \mT
    &= \gamma_1\left(
          \frac{\mF^-}{\alpha^2} - \frac{1}{2}(\mF^0)^2 - \partial \mF^0
          - \frac{1}{2\alpha^4} \left( \mF^N \right)^2
        \right)
      + 2\gamma_2\left(
          \mF^- - \mF^N \mF^0 - \partial \mF^N
        \right),\\
  \mJ
    &= \left( 2\gamma_2 - \frac{\gamma_1}{\alpha^2} \right)
      \left( \mF^0 - \frac{\mF^N}{\alpha^2} \right).
\end{align}
Their variation is given by
\begin{align}
	\delta \mT \label{eq:finite-alpha-kappaP-variation}
		&= \lambda^+ \partial \mT + 2\mT\partial\lambda^+
			+ \gamma_1 \partial^{3} \lambda^+
			+ \mJ \partial \bar\lambda^N
			- 2\left(\gamma_2 - \frac{\gamma_1}{2\alpha^2} \right) \partial^{2}\bar\lambda^N,\\
  \delta \mJ \label{eq:finite-alpha-kappaN-variation}
    &= \lambda^+ \partial \mJ + \mJ \partial\lambda^+
      + \left( 2\gamma_2 - \frac{\gamma_1}{\alpha^2} \right)
      \left(\partial^{2}\lambda^+
      + \frac{2\partial\bar\lambda^N}{\alpha^2} \right).
\end{align}
As before we can write the charge algebra in terms of the Fourier modes
\begin{equation}
  \LL_n = - \oint d\vphi\, e^{in\vphi} \mT(\xp), \qquad
	\mN_n = - \oint d\vphi\, e^{in\vphi} \mJ(\xp).
\end{equation}
They satisfy the following commutation relations,
\begin{subequations}
		\label{eq:P11-finite-alpha-ASA}
\begin{align}
  [\LL_m, \LL_n]
		&= (m-n) \LL_{m+n} + 2\pi \gamma_1 m^3 \delta_{m+n,0}, \\
  [\LL_m, \mN_n]
  &= - n \mN_{m+n}
    - 2\pi i \, m^2
      \left( 2\gamma_2 - \frac{\gamma_1}{\alpha^2} \right)
      \delta_{m+n,0},\\
  [\mN_m, \mN_n]
		&= - \frac{4\pi m}{\alpha^2}
			\left( 2\gamma_2 - \frac{\gamma_1}{\alpha^2} \right)
				\delta_{m+n,0}.
\end{align}
\end{subequations}
This is a Virasoro algebra with \emph{coupled} affine $u(1)$ algebra and nonzero affine $u(1)$ level.
The contracted algebra can be obtained by sending $\alpha\to\infty$, and we indeed reproduce the asymptotic symmetry algebra at infinite $\alpha$ in \eqref{eq:P11-infinite-alpha-ASA}.

\subsection{Relation to uncoupled algebra}\label{ssec:sectionuncouple}
This result may seem confusing.
One would expect that the asymptotic symmetry algebra of an $sl(2,\RR)\oplus u(1)$ Chern--Simons theory would be an uncoupled Virasoro and affine $u(1)$ algebra.
In fact, we can easily include a $u(1)$ generator $N_0$ in the computation of appendix \ref{app:cs-asymptotics-sl2-u1-decoupled} and show that this is the case.
The constrained connection, residual gauge transformations and boundary charges are then
\begin{eqnarray}
  a_+ &=& L_+ + F^0 L_0 + F^- L_- + F^N N_0, \\
  \lambda
    &=&  \lambda^+ L_+
      + (F^0 \lambda^+ - \partial \lambda^+) L_0
      + \lambda^- L_-
      + \lambda^N N_0, \\
	Q_\lambda
    &=& -2 \oint \langle \lambda, \delta a_+ \rangle
		= \oint d\varphi\,
			\left(
				\lambda^+ T + \lambda^N J
			\right)
		= Q^\text{Vir}[\lambda^+]
			+ Q^{u(1)}[\lambda^N].
\end{eqnarray}
Here, we have defined the physical currents
\begin{align}
	T
		&= 2\gamma_s
      \left(
  		  F^- - \frac{1}{4} (F^0)^2 - \frac{1}{2} \partial F^0
			\right),
    \qquad
  J
    = - \gamma_u F^N\,.
\end{align}
Indeed, they transform as Virasoro and affine $u(1)$ currents,
\begin{align}
	\delta T
		&= \lambda^+ \partial T + 2 T \partial \lambda^+
			+ \gamma_s \partial^{3} \lambda^+,
		\qquad
	\delta J
		= - \gamma_u \partial \lambda^N.
\end{align}
Again, we can decompose these currents in Fourier modes,
\begin{align}
	L_n &= - \oint d\varphi\, e^{in\varphi}\, T(\xp),
    \qquad
	N_n = - \oint d\varphi\, e^{in\varphi} J(\xp).
\end{align}
These charges satisfy an \emph{uncoupled} Virasoro-affine $u(1)$ algebra with nonzero affine level,
\begin{subequations}
  \begin{align}
  	[L_m, L_n]
  		&= (m-n) L_{m+n}
  			+ 2\pi\gamma_s m^3 \delta_{m+n,0}, \\
    [L_m, N_n]
      &= 0,\\
  	[N_m, N_n]
  		&= 2\pi \gamma_u \, m\, \delta_{m+n,0}.
  \end{align}
\end{subequations}

In fact, there is no contradiction here.
The uncoupled symmetries can be transformed into the coupled algebra at finite $\alpha$ in \eqref{eq:P11-finite-alpha-ASA}.
It is easiest to see this on the level of the current transformations.
The first step is to match $\delta J$ with $\delta\mJ$.
For this, define
\begin{equation}
  - \gamma_u \lambda^N
    = \lambda^+ J
      - \frac{2\gamma_u}{\alpha^2} \left( \partial\lambda^+ + \frac{2\bar\lambda^N}{\alpha^2} \right).
\end{equation}
Then $\delta J$ reproduces \eqref{eq:finite-alpha-kappaN-variation}.
To obtain the correct Virasoro transformations, define
\begin{equation}\label{eq:uncoupled-coupled-virasoro-current-redefinition}
  \mT
    = T - \frac{\alpha^4}{8\gamma_u} J^2
      - \frac{\alpha^2}{2} \partial J.
\end{equation}
This current satisfies the coupled transformation relation \eqref{eq:finite-alpha-kappaP-variation}.
The identification between the coupled and uncoupled modes $(\LL_n,\mN_n)$ and $(L_n,N_n)$ then follows by expanding the above.
As generators of symmetries on a classical phase space, the coupled and uncoupled algebras are therefore equivalent.

\section{Discussion and Outlook}
\label{sec:discussions}

We conclude with a  discussion and perspectives for further work.

The results of this paper show that many
of the features of the AdS$_3$/CFT$_2$ correspondence can be realized in a novel holographic correspondence, involving a pseudo-Newton--Cartan theory in the bulk and a particular near-BPS limit on the boundary.
This provides a concrete model of beyond-AdS/CFT holography,
opening up many avenues of further exploration in terms of generalizing other well-studied aspects of AdS/CFT.

As one possible direction, we note that the two copies of the Virasoro spacetime algebra in AdS$_3$/CFT$_2$ can be induced from a $sl(2,\RR)\oplus sl(2,\RR)$ current algebra on the string worldsheet \cite{Giveon:1998ns}.
Here, the spacetime chirality is closely related to worldsheet chirality. For example, the  left moving chiral algebra in spacetime is lifted from left movers on the string worldsheet and vice versa.
Moreover, the string worldsheet analysis can give a microscopic interpretation of the central charge in terms of string winding modes. It is expected that similar string theory analyses should be valid even after we take the In\"on\"u-Wigner limit.
One would expect the corresponding world-sheet theory to  be a WZW model on $P_2^c$, similar to what is studied in \cite{Nappi:1993ie,Kiritsis:1993jk}.
In connection to this, it is interesting to note that the target space of such a model can easily be inferred from the results of \cite{Nappi:1993ie} by replacing $E_2^c$ with $P_2^c$.
This leads precisely to a
pp-wave like geometry with signature  $(-1,-1,1,1)$, i.e.  two `times', such that after a null reduction one obtains a pseudo-NC geometry in the same way that pp-waves connect to NC geometry after null reduction \cite{Christensen:2013rfa,Duval:1984cj,Duval:1990hj,Julia:1994bs}.

Moreover, it would be interesting to study string theory on pseudo-Newton--Cartan gravity,  using the
AdS$_3$/CFT$_2$ results of Ref. \cite{Giveon:1998ns}, while one could also examine
whether the Wakimoto representation of $sl(2,\RR)\oplus u(1)$ on the worldsheet \cite{Giveon:1999jg} can reproduce the representations of \cite{Kiritsis:1993jk} by taking our contraction limit.
Moreover, in a recent work \cite{Harmark:2017rpg} it was shown that NC geometry appears as the target space
in nonrelativistic string theory,  which may also be of use to understand strings on pseudo-NC geometry.
In connection to this, it also seems relevant to note that the nonrelativistic limit of AdS/CFT considered in
\cite{Gomis:2005pg} shows that the resulting nonrelativistic string
action has the supersymmetric Newton--Hooke group as a symmetry group.

Another worthwhile direction to pursue is to employ the concrete AdS/CFT model coming from the D1-D5 brane system in type IIB superstring theory, which provides a duality between $\mN=(4,4)$ superconformal field theory and string theory in AdS$_3$ \cite{Maldacena:1997re,Giveon:1998ns,David:1999nr}.
Thus our gravity theory should have a supersymmetric extension, which is related to an appropriate limit of $\mN=(2,2)$ supergravity in three dimensions \cite{Izquierdo:1994jz}.
In fact, the bosonic sector of this supergravity theory exactly has two $u(1)$ gauge fields as R-charge currents so directly fits into the symmetry algebra we took as our starting point.
Understanding this string theory and supergravity embedding after our limit should provide a rich structure as well.

At the level of solutions of pseudo-NC gravity, we have focused on the vacuum but an obvious next step is to examine the limit of the BTZ black hole  \cite{Banados:1992wn} and its physics.
Another class of solutions
that could shed further light on the theory are  the BPS supergravity solutions of \cite{Maldacena:2000dr,Balasubramanian:2000rt,Lunin:2002iz,Lunin:2001jy,Lunin:2002bj,Kanitscheider:2006zf,Skenderis:2006ah,Kanitscheider:2007wq} which are dual to CFT chiral primaries.
More generally, one may wish to address bulk reconstruction in our setup.
While for AdS$_3$/CFT$_2$ the entire relativistic bulk should be reconstructed from the boundary conformal field theory,
pseudo-NC gravity represents in some sense a more minimal setup.
In this case, we only need to reconstruct a foliation structure, namely two-dimensional pseudo-Riemannian geometry fibered over a dilatation one-form.
Another, related direction would be to see if there is an analogue of holographic entanglement entropy
\cite{Ryu:2006bv} for our correspondence.
For this, a minimal setup can be constructed using Chern--Simons theory \cite{Ammon:2013hba,deBoer:2013vca}.
It would also be very interesting to investigate the implications of the radial fibration on the RG flow of the dual field theories.

More generally, it will be important to better understand the field theory that is dual to pseudo-NC gravity.
In this connection it is worth remarking that our limit has a strong resemblance to the limit that gives rise to Spin Matrix Theory \cite{Harmark:2014mpa}, which follows from the correspondence between AdS$_5$/CFT$_4$.
To see this, define a coupling
constant $g=\alpha^{-2}$ and  identify the energy $E$ and charge $J$ as $E=\alpha^{-2}D=\mathcal{N}+\frac{g}{2}\mathcal{D}$ and $J=-\alpha^{-2}Q_1=\mathcal{N}-\frac{g}{2}\mathcal{D}$ respectively.
By the state-operator map the dilatation operator corresponds to the energy of states of the theory on the cylinder.
In Spin Matrix Theory, $\mathcal{N}$ is the length of the spin chain and $\mathcal{D}$ is the one-loop dilatation operator of the spin chain.
The limit $\alpha\rightarrow\infty$ zooms in on states close to $E=J$ which is the lowest lying state in the spectrum, so by unitarity we have $E\ge J$.
The Spin Matrix Theory limit \cite{Harmark:2014mpa} corresponds to sending $g\to 0$ while keeping $(E-J)/g$ fixed, which is precisely what happens in our limit for the operators $D$ and $Q_1$ as we take $\alpha\to\infty$.
This connection with Spin Matrix Theory seems to suggest that the symmetry of Spin Matrix Theory might be related to an another real form of the complexified Newton--Hooke algebra.
On the other hand, applied to AdS$_3$/CFT$_2$, especially in relation to the $\mN=(4,4)$ superconformal field theory and its possible spin chain interpretation (see \cite{Baggio:2017kza} for recent progress), this suggests that there might be some form of two-dimensional Spin Matrix Theory.

Regarding the field theory interpretation,  it is important to emphasize that the finite $\alpha$ identifications we
have made in Section \ref{ssec:sectionuncouple} are  entirely classical.
We have found an infinite-dimensional algebra of conserved charges for classical Chern--Simons theory, which correspond to real-valued functions on the phase space.
Upon quantization, these charges should lead to unitary operators on a Hilbert space of states.
However, our classical computations do not tell us what the inner product or Hermitian conjugate on this Hilbert space should be.
We have only obtained the algebra of symmetries that should be realized on it.
Unitary representations of the coupled algebra do exist, and \cite{Afshar:2015wjm} has explained how to construct induced representations using its semidirect product structure.
It would be very interesting to study their consequences from a field theory perspective.
While the bulk may have allowed us to identify the relevant symmetries, we believe that field theory will be our guide towards their representations.

Finally, it could be interesting to find analogues of our holographic correspondence, such as higher spin and/or higher-dimensional generalizations.
Following the higher spin AdS${}_3$ work of e.g. \cite{Gaberdiel:2010pz}, and its Chern--Simons theory construction \cite{Campoleoni:2010zq,Henneaux:2010xg}, it is possible to find connections between the non-AdS higher spin holography of  \cite{Gary:2012ms,Afshar:2012nk}, TMG holography  \cite{Anninos:2008fx,Grumiller:2008es,Grumiller:2008pr,Li:2008dq,Chen:2011yx,Chen:2011vp} and
nonrelativistic higher spin works of \cite{Bergshoeff:2016soe} to the higher spin generalization of our work.

\subsubsection*{Acknowledgments}

We thank Marieke van Beest, Jan de Boer, Alejandra Castro, Troels Harmark and Diego Hofman for useful discussions.
The work of NO is supported in part by the project ``Towards a deeper understanding of black holes with nonrelativistic holography'' of the Independent Research Fund Denmark
(grant number DFF-6108-00340).
JH, YL and GO acknowledge hospitality of the Niels Bohr Institute and NO acknowledges hospitality of University of Amsterdam during part of this work.
The  work  of  JH  was supported in  part by  a  STSM Grant  from COST Action MP1405 QSPACE.
YL thanks Kyung Hee University for hospitality where part of this work was presented.
YL also acknowledges support from Thousand Young Talents Program of Prof.~Wei~Li.
GO is supported by the Foundation for Fundamental Research on Matter (NWO-I) and is especially thankful to Diego Hofman for his advice and collaboration on related projects.

\appendix

\section{Review of \boldmath\texorpdfstring{$so(2,2)$}{so(2,2)} Chern--Simons theory}
\label{app:so22-cs-review}
Here we review some facts from the standard $so(2,2)$ relativistic AdS$_3$ Chern--Simons theory that will be of use in the main text.
In particular, we recall the identification with Einstein gravity with negative cosmological constant and find the connection corresponding to the global AdS$_3$ metric.
We also see how the fact that $so(2,2)$ splits in two copies of $sl(2,\RR)$ allows for a simple way to deal with the boundary terms in the variational problem of the Chern--Simons action.

\subsection{Connection and action}
The $so(2,2)$ Chern--Simons (CS) connection consists of the vielbein and spin connection,
\begin{equation}\label{eq:app-so22-connection-split}
	\mathbf{A} = E^A T_A + \Omega^A J_A
		= A^A S_A + \bar A^A\bar S_A\,,
	\quad
	A^A = \Omega^A + \frac{1}{l} E^A\,,
	\quad
	\bar A^A = \Omega^A - \frac{1}{l} E^A\,.
\end{equation}
where $S_A, \bar{S}_A$ are $so(2,1)$ generators.
In terms of the $so(2,2)$ generators in \eqref{eq:app-so(2,2)-alg},
\begin{equation}\label{eq:StoJT}
S_A=\frac{1}{2}\left(J_A+ lT_A\right)\,,\qquad\bar S_A=\frac{1}{2}\left(J_A-lT_A\right).
\end{equation}
We work with Hermitian generators, which satisfy the commutation relations
\begin{equation}\label{eq:app-so(2,2)-to-so(2,1)}
\left[ S_A, S_B \right]
= i\tensor{\epsilon}{_{AB}^C} S_C\,,\qquad
\left[\bar S_A,\bar S_B \right]
= i\tensor{\epsilon}{_{AB}^C}\bar S_C.
\end{equation}
The invariant Killing metric is given in the $S_A$ basis by
\begin{equation}
\langle S_A, S_B \rangle
= \frac{1}{2} \gamma_s \eta_{AB},
\quad
\langle \bar S_A, \bar S_B \rangle
=- \frac{1}{2}\bar\gamma_s \eta_{AB},
\quad
\langle S_A, \bar S_B \rangle = 0\,.
\end{equation}
Here, $\gamma_s$ is an arbitrary real constant.\footnote{
  For $so(2,1)\simeq sl(2,\RR)$, it is usually parametrized by $k/4\pi$.
  }
With this decomposition, the $so(2,2)$ Chern--Simons Lagrangian density can be split into two $so(2,1)$ factors
\begin{eqnarray}
	\mathcal{L}_\text{CS}
		&=&
      \left\langle\mathbf{A}, d\mathbf{A}
        - \frac{2i}{3}\mathbf{A} \wedge\mathbf{A}
      \right\rangle
		= \mathcal{L}_\text{CS}[A] + \mathcal{L}_\text{CS}[\bar A]\,,\nonumber \\
	\mathcal{L}_\text{CS}[A]
		&=&
			\frac{1}{2}\gamma_s \left(
				\eta_{AB} A^A \wedge d A^B
					+ \frac{1}{3} \epsilon_{ABC} A^A \wedge A^B \wedge A^C
			\right).
\end{eqnarray}
The factor of $-i$ in the CS action on the right hand side of the first equality is due to the fact that we work with Hermitian generators.
In terms of the metric data, this gives
\begin{align}
	\mathcal{L}_\text{CS}
		&= \frac{\gamma_s +\bar\gamma_s}{2l}
			\left(
						2 E^A \wedge d \Omega^B \eta_{AB}
						+\epsilon_{ABC}  E^A \wedge\Omega^B \wedge \Omega^C
				+ \frac{1}{3l^2}
					\epsilon_{ABC} E^A \wedge E^B \wedge E^C
			\right)\nonumber\\
		&{}\qquad
      + \frac{\gamma_s -\bar\gamma_s}{2}
			\left(
						\Omega^A \wedge d \Omega^B \eta_{AB}
						+ \frac{1}{3} \epsilon_{ABC}\Omega^A \wedge \Omega^B \wedge \Omega^C+ \frac{1}{l^2}E^A \wedge d E^B \eta_{AB}\right.\nonumber\\
    &{}\qquad\left.
      + \frac{1}{l^2} \epsilon_{ABC}
        E^A \wedge E^B \wedge \Omega^C\right).
      \label{eq:app-general-so22-cs-action}
\end{align}
The term proportional to $\gamma_s -\bar\gamma_s$ is the Einstein-Hilbert Lagrangian density with negative cosmological constant $\Lambda=-{1}/{l^2}$.
The term proportional $\gamma_s -\bar\gamma_s$ is the Lorentz--Chern--Simons term.
In this paper, we will only consider the Einstein term and set $\gamma_s = \bar\gamma_s$.

\subsection{Global \texorpdfstring{AdS$_3$}{AdS3} solution}
\label{app:global-ads3-sol}
The curvature of a Hermitian connection $\mathbf{A}$ is given by
\begin{equation}
  \mathbf{F} = d\mathbf{A} - i \mathbf{A} \wedge \mathbf{A}.
\end{equation}
Its $J_A$ and $T_A$ components give the torsion-free condition and the 3D Einstein equation,
\begin{align}
	0 &= \label{eq:app-so22-eom-torsion}
		dE^A + \tensor{\epsilon}{^A_{BC}} \Omega^B\wedge E^C, \\
	0 &= \label{eq:app-so22-eom-flatness}
		d\Omega^A + \frac{1}{2}\tensor{\epsilon}{^A_{BC}}\left(
				\Omega^B\wedge\Omega^C
				+\frac{1}{l^2} E^B\wedge E^C
			\right).
\end{align}
For a given metric, we can solve the spin connection $\Omega^A$ in terms of the corresponding vielbein $E^A$ using the torsion-free condition.
The second equation is then a constraint on the curvature of this spin connection.
In the case of global AdS$_3$, we can take
\begin{gather}
	ds^2 = \ell^2 \left(
			- \cosh^2\rho dt^2 + \sinh^2\rho d\vphi^2 + d\rho^2
		\right), \nonumber\\
	E^0 = \ell \cosh\rho dt, \quad
	E^1 = -\ell \sinh\rho d\vphi, \quad
	E^2 = \ell d\rho.
\end{gather}
Solving \eqref{eq:app-so22-eom-torsion} then leads to
\begin{equation}
	\Omega^0 =  \cosh\rho d\vphi, \quad
	\Omega^1 =  -\sinh\rho dt, \quad
	\Omega^2 = 0.
\end{equation}
Following \eqref{eq:app-so22-connection-split}, the corresponding $so(2,1)$ connections are
\begin{align}
	A	&=  2\cosh\rho\, S_0 d\xp - 2\sinh\rho\, S_1 d\xp
					+ S_2 d\rho, \label{eq:Aglobal}\\
	\bar A &= 2 \cosh\rho\, \bar S_0 d\xm + 2\sinh\rho\, \bar S_1 d\xm
					-\bar S_2 d\rho.
\end{align}
Here we have introduced the null coordinates $x^\pm =\frac{1}{2}( \vphi \pm t)$. In Fefferman--Graham coordinates with $r=e^\rho$ the global AdS$_3$ metric reads
\begin{equation}
ds^2=\frac{dr^2}{r^2}-\left(r^2+2+r^{-2}\right)\frac{1}{4}dt^2+\left(r^2-2+r^{-2}\right)\frac{1}{4}d\varphi^2\,.
\end{equation}

Now let us move on to the $sl(2,\RR)$ basis.
We define $L_{-1}$, $L_0$, $L_1$ as
\begin{equation}\label{eq:L}
L_{-1}=S_0+S_1\,,\qquad L_0=S_2\,,\qquad L_1=S_0-S_1\,.
\end{equation}
Likewise we define $\bar L_{-1}$, $\bar L_0$, $\bar L_1$ as
\begin{equation}\label{eq:barL}
\bar L_{-1}=-(\bar S_0-\bar S_1)\,,\qquad\bar L_0=-\bar S_2\,,\qquad\bar L_1=-(\bar S_0+\bar S_1)\,.
\end{equation}
The $L_m$ and $\bar L_m$ generators with $m,n=-1,0,1$ both satisfy the $sl(2,\RR)$ algebra
\begin{equation}
[L_m,L_n]=i(m-n)L_{m+n}\,.
\end{equation}

Using the definitions \eqref{eq:L} and \eqref{eq:barL}, the global AdS$_3$ connection \eqref{eq:Aglobal} becomes
\begin{align}
	A &= L_0 d\rho
					+\left(
						 e^{\rho}\, L_1+e^{-\rho}\, L_{-1}
					\right)d\xp
					= e^{i \rho L_0} \left(
						id + (L_1+L_{-1}) d\xp
					\right) e^{-i\rho L_0}, \\
	\bar A &=  \bar L_0 d\rho
					-  \left(
						e^\rho\, \bar L_1 + e^{-\rho}\, \bar L_{-1}
					\right)d\xm
					= e^{i\rho \bar L_0} \left(
						id -(\bar L_1 + \bar L_{-1}) d\xm
					\right) e^{-i\rho \bar L_0}.\nonumber
\end{align}
Since the two $sl(2,\RR)$ sectors commute, we can write this concisely as
\begin{equation}\label{eq:app-AdS3-background-connection}
	\mathbf{A}_{AdS_3}
		= e^{i\rho(L_0 + \bar L_0)} \left(
				id +(L_1+L_{-1}) d\xp - (\bar L_1+ \bar L_{-1}) d\xm
			\right)e^{-i\rho(L_0 + \bar L_0)}.
\end{equation}
This is the $so(2,2)$ connection corresponding to a global AdS$_3$ background.

\subsection{Variational problem}\label{app:variational-problem}
In the above, we found flat connections by solving $\mathbf{F}=0$.
However, we still need to do some work to show that this is actually the equation of motion of Chern--Simons theory.
This discussion is not particular to $so(2,2)$, so in this subsection we will use $A$ to denote a general connection with curvature $F$, both valued in a Lie algebra $\mathfrak{g}$.
The problem is that the variation of the Chern--Simons action
\begin{equation}\label{eq:cs-variational-problem}
	\delta S_\text{CS}
			= 2 \int_{M} \left\langle\delta A, F\right\rangle
				+ \int_{\partial M}
					\left\langle A, \delta A\right\rangle
\end{equation}
can be nonzero around flat connections on a manifold $M$ with boundary.
Thus we cannot claim that $F=0$ is the equation of motion unless the boundary term vanishes.
There are various ways to achieve this, and we will choose the simplest solution.
More complicated solutions allow us to construct a boundary phase space with nonzero chemical potentials.

Let us choose coordinates $x^\pm=\vphi\pm t$ on the boundary $\partial M$, and parametrize the transverse direction by $\rho$.
We can expand the connection $A$ as follows,
\begin{equation}
  A = A_\rho d\rho + A_- d\xm + A_+ d\xp.
\end{equation}
Using a gauge transformation, $A_\rho$ can be set to an arbitrary constant (see for example  \cite{Campoleoni:2010zq}).
The simplest way to obtain a well-defined variational problem is to make sure that the boundary term in \eqref{eq:cs-variational-problem} vanishes.
We can do this by requiring
\begin{equation}\label{eq:cs-variational-problem-Amin-zero-sol}
	\left.A_-\right|_{\partial M} = 0
		\quad \implies \quad
	\left\langle A, \delta A\right\rangle
		= \left(
        \left\langle A_+ , \delta A_-\right\rangle
        - \left\langle A_- , \delta A_+ \right\rangle
      \right)
      d\xp\wedge d\xm
		= 0.
\end{equation}
With such a boundary condition, $F=0$ is a well-defined equation of motion.
In fact, $F_{\rho-}=0$ propagates the constraint $A_-=0$ on the boundary all the way into the bulk.
We then end up with a connection of the form
\begin{equation}
  A = b(\rho)\inv d b(\rho) + b(\rho)\inv a(\xp) b(\rho),
    \quad
  a = J^a(\xp) T_a d\xp.
\end{equation}
The currents $J^a$ parametrize a Wess-Zumino-Witten affine $\mathfrak{g}$-phase space on the boundary $\partial M$, where $\mathfrak{g}$ the same Lie algebra as was used to write down the Chern--Simons algebra.

\paragraph{Split algebra and chirality}
While \eqref{eq:cs-variational-problem-Amin-zero-sol} solves the variational problem, we do not want to set the $d\xm$ component to zero for the entire $so(2,2)$ connection $\mathbf{A}$.
As we can see from \eqref{eq:app-so22-connection-split}, this would mean that the vielbein $E^A$ cannot contain $d\xm$ components so the resulting metric is degenerate in the $\xm$ coordinate, which is clearly undesirable.

However, if the algebra in which $\mathbf{A}$ is valued splits, one can impose the above boundary conditions and at the same time have a nondegenerate metric.
Returning to $so(2,2)$, using
\begin{equation}
	so(2,2)
    \simeq sl(2,\RR) \oplus sl(2,\RR),
\end{equation}
we can demand that the $d\xp$ respectively the $d\xm$ component of either factor vanishes.
This choice is compatible with the AdS$_3$ background connection \eqref{eq:app-AdS3-background-connection}.
In other words, we choose the following chirality for the $sl(2,\RR) \oplus sl(2,\RR)$ connection,
\begin{equation}
  \label{eq:app-chirality-choice}
  L_a d\xp, \qquad
  \bar L_a d\xm.
\end{equation}

\subsection{General flat connections and their metric data}
Furthermore, the background connection \eqref{eq:app-AdS3-background-connection} also motivates us to set the $so(2,2)$ coefficients of the radial component to
\begin{equation}
  \mathbf{A}_\rho
    = L_0 + \bar L_0.
\end{equation}
With the chirality choice in \eqref{eq:app-chirality-choice}, we can then write the \emph{most general} flat connection as
\begin{equation}\label{eq:app-general-so22-connection-currents}
	\mathbf{A} = e^{i\rho(L_0 +\bar L_0)} \left(
				id + F^a(\xp) L_a d\xp + \bar{F}^a(\xm) \bar L_a\, d\xm
			\right)e^{-i\rho(L_0 +\bar L_0)}.
\end{equation}
Now let us expand \eqref{eq:app-general-so22-connection-currents} and write it in terms of the $so(2,1)$ generators $S_A$ and $\bar S_A$ in \eqref{eq:StoJT},
\begin{align}
	\mathbf{A} &= \left(L_0 +\bar L_0 \right) d\rho
			+ \left(
					e^{\rho} F^+ L_1 + F^0 L_0 + e^{-\rho} F^-L_{-1}
			\right)d\xp\nonumber\\
	&{}\qquad + \left(
			e^{\rho} \bar F^+ \bar L_1+ \bar F^0 \bar L_0 + e^{-\rho} \bar F^- \bar L_{-1}
		\right)d\xm \nonumber \\
	&=  \left(d\rho + F^0 d\xp\right)S_2
		+ \left(- d\rho -\bar F_0 d\xm\right)\bar S_2 \\
	&{}\qquad
		+\left(
			e^{\rho}F^+ d\xp + e^{-\rho} F^- d\xp
		\right) S_0
		+\left(-
			e^{\rho}F^+ d\xp + e^{-\rho} F^- d\xp
		\right) S_1 \nonumber \\
	&{}\qquad
		+\left(
			-e^{\rho}\bar F^+ d\xm - e^{-\rho}\bar F^-d\xm
		\right)\bar S_0
		+\left(
			-e^{\rho}\bar F^+ d\xm + e^{-\rho} \bar F^- d\xm
		\right)\bar S_1. \nonumber
\end{align}
We can then use the relation to $J_A$ and $P_A$ in \eqref{eq:StoJT} to write down the vielbein and spin connection corresponding to the flat connection \eqref{eq:app-general-so22-connection-currents},
\begin{subequations}
	\label{eq:app-general-so22-vielbein-spin-connection-data}
\begin{align}
	E^0 &= \frac{l}{2}
		\left(
			e^\rho \left( F^+ d\xp +\bar F^+ d\xm \right)
			+ e^{-\rho} \left( F^- d\xp + \bar F^- d\xm \right)
		\right), \\
	E^1 &= \frac{l}{2}
		\left(
			- e^\rho \left( F^+ d\xp -\bar F^+ d\xm \right)
			+ e^{-\rho} \left( F^- d\xp -\bar F^- d\xm \right)
		\right), \\
		E^2 &= l
		\left(
			 d\rho + \frac{1}{2}F^0 d\xp +\frac{1}{2}\bar F^0 d\xm
		\right), \\
	\Omega^0 &=\frac{1}{2}
		\left(
			e^\rho \left( F^+ d\xp -\bar F^+ d\xm \right)
			+ e^{-\rho} \left( F^- d\xp - \bar F^- d\xm \right)
		\right),\\
	\Omega^1 &=  \frac{1}{2}
		\left(
			- e^\rho \left( F^+ d\xp +\bar F^+ d\xm \right)
			+ e^{-\rho} \left( F^- d\xp +\bar F^- d\xm \right)
		\right),\\
			\Omega^2 &= \frac{1}{2}
		\left(
			F^0 d\xp -\bar F^0 d\xm
		\right).
\end{align}
\end{subequations}
Requiring that such a connection agrees with the AdS$_3$ connection \eqref{eq:app-AdS3-background-connection} at leading radial order leads to the constraints
\begin{equation}\label{eq:app-aads3-constraint}
	F^+ \equiv 1, \quad \overline{F}^+ \equiv - 1.
\end{equation}
As we will see in the following, these constraints reduce the boundary WZW model to the Brown-Henneaux asymptotic Virasoro symmetries.

\section{Review of Chern--Simons asymptotic symmetries  \label{sec:CSasymptsym}}
\label{app:cs-asymptotics-sl2-u1-decoupled}
The purpose of this section is to give a detailed review of the reduction of the boundary Wess-Zumino-Witten model of $sl(2,\RR)$ Chern--Simons under the AdS$_3$ constraint \eqref{eq:app-aads3-constraint}.
This procedure, which is also known as Drinfeld-Sokolov reduction, will produce a Virasoro algebra of asymptotic charges, which is (one chiral half of) the usual Brown-Henneaux asymptotic symmetry algebra.

\subsection*{Gauge transformations}
We work with the $sl(2,\RR)$ generators $L_a$ defined in \eqref{eq:sl2-sl2bar-commutators}.
Following \eqref{eq:app-general-so22-connection-currents}, we can write the most general connection for a single $sl(2,\RR)$ factor as
\begin{equation}
  \label{eq:app-general-chiral-connection-standard-form}
  A = b \left( id + a_+ d\xp \right) b\inv.
\end{equation}
Here, we have introduced two $sl(2,\RR)$-valued functions
\begin{equation}
  b(\rho) = e^{i\rho L_0}, \quad
  a_+(\xp) = F^a(\xp) L_a.
\end{equation}
The curvature $F=dA-iA\wedge A$ transforms covariantly with respect to the gauge transformation $\delta_\Lambda A=d\Lambda-i[A,\Lambda]$.
The form \eqref{eq:app-general-chiral-connection-standard-form} is preserved by gauge transformations of the form $\Lambda=b\lambda (\xp) b^{-1}$.
Under such a gauge transformation, the reduced connection $a$ transforms as
\begin{equation}
  \delta_\lambda a=d\lambda-i[a,\lambda].
\end{equation}
The fact that the residual symmetries of connections of the form \eqref{eq:app-general-chiral-connection-standard-form} take such an easy form simplifies our analysis.
Once we have agreed on a choice of $A_\rho$, we only have to concern ourselves with the reduced connection $a_+(\xp)d\xp$ and its symmetries $\lambda(\xp)$.
In fact, these transformations are symmetries only up to boundary terms, which lead to boundary charges
\begin{equation}\label{eq:app-cs-boundary-charges}
  \delta Q_\lambda
  = -2 \oint_{\partial\Sigma} \langle \lambda, \delta_\lambda a \rangle.
\end{equation}
If they are integrable, these boundary charges satisfy the following Poisson bracket
\begin{equation}
  \{ Q_\lambda, Q_\mu\}
    = \delta_\lambda Q_\mu
    = -2 \oint_{\partial \Sigma} \langle \mu, \delta_\lambda a \rangle
    = -i Q_{[\lambda,\mu]} + 2 \oint_{\partial \Sigma} \langle \lambda, d\mu\rangle.
\end{equation}
For a general Lie algebra with generators $T_a$, let us denote the invariant bilinear metric by $\Omega_{ab} = \langle T_a, T_b\rangle$.
We only consider nondegenerate bilinear forms, whose inverse exists and is denoted by $\Omega^{ab}$.
Furthermore, we define real structure constants $\tensor{f}{_{ab}^c}$ using $[T_a,T_b] = i \tensor{f}{_{ab}^c}T_c$.
With these conventions, the Poisson bracket on the currents $F^a(\xp)$ is given by
\begin{equation}\label{eq:app-current-poisson-bracket}
  \{F^a(\xp), F^b(y^+)\}
    = - \frac{1}{2} \tensor{f}{^{ab}_c} F^c(\xp) \delta (\xp-y^+)
      + \frac{1}{2} \partial_{\xp} \delta (\xp-y^+) \Omega^{ab}.
\end{equation}

\subsection*{Constraint}
Imposing the asymptotically AdS$_3$ constraint \eqref{eq:app-aads3-constraint}, we restrict ourselves to
\begin{equation}\label{eq:bdrycondition}
	a_+ = L_+ + F^0 L_0 + F^- L_-.
\end{equation}
The most general residual transformations that preserves this form is given by
\begin{equation}\label{eq:sl2-ds-parameter-constr}
	\lambda = \lambda^a L_a, \quad
	\lambda^0 = -\partial \lambda^+ + \lambda^+ F^0.
\end{equation}
In the following, we will see that $\lambda^+$ is the transformation parameter for a Virasoro algebra of boundary charges.
Now we must find the corresponding current.

\subsection*{Invariant polynomials}
We imposed a constraint but did not gauge fix the corresponding degrees of freedom.
The remaining degrees of freedom in the connection \eqref{eq:bdrycondition} therefore only correspond to \emph{one} physical current.
One way to understand what the building blocks are that make up the physical current is as follows.

The constraint $F^+\equiv1$ generates gauge transformations in the phase space \eqref{eq:bdrycondition}, which is parametrized by the functions $F^0$ and $F^-$.
Using the Poisson bracket \eqref{eq:app-current-poisson-bracket}, we see that these functions transform under the constraint gauge transformations as follows,
\begin{align}
	\delta_\Lambda F^0(x)
		&= \int dy \left\{ \Lambda(y) (F^+(y) - 1), F^0(x) \right\}
		= \frac{1}{\gamma_s} \Lambda(x) F^+(x)
		\equiv \frac{1}{\gamma_s} \Lambda(x), \\
	\delta_\Lambda F^-(x)
		&= \int dy \left\{ \Lambda(y) (F^+(y) - 1), F^-(x) \right\}
		= \frac{1}{2\gamma_s} \left(
				 \Lambda(x) F^0(x)
				+ \partial_x \Lambda(x)
			\right).
\end{align}
This shows us how to make an invariant combination out of these current components that parametrizes the physical current on the constrained phase space,
\begin{equation}\label{eq:sl2-invar-current}
	F^\text{inv}
		= F^- - \frac{1}{4}( F^0 )^2 -  \frac{1}{2} \partial F^0,
  \quad
  \delta_\Lambda F^\text{inv} = 0.
\end{equation}
We will see below that this combination transforms as a Virasoro current.

\subsection*{Charge integrand and current transformations}
The infinitesimal boundary charges can be obtained as follows.
We will see that they can be expressed in terms of the invariant current we found above.
Integrating over constant time slices, we find that the usual expression for Chern--Simons boundary charges evaluates to
\begin{align}
	\delta Q_\lambda
	 		&= -2\oint d\varphi\,
				\langle \lambda, \delta a_+ \rangle\nonumber \\
			&= -2\gamma_s \oint d\varphi\,
				\left(
					- \lambda^+ \delta F^-
					+ \frac{1}{2} \lambda^0 \delta F^0
				\right)
          \\
			&= -2\gamma_s \oint d\varphi\,
				\left(
					- \lambda^+ \delta F^-
					+ \frac{1}{2} \left(-\partial \lambda^+ + \lambda^+ F^0\right)
						\delta F^0
				\right)
          \\
			&= \oint d\varphi\,
					\lambda^+ \delta T.
        \nonumber
\end{align}
This corresponds to a charge with parameter $\lambda^+$ and current
\begin{equation}
	T
		= 2\gamma_s \left(
				F^- - \frac{1}{4} (F^0)^2 - \frac{1}{2} \partial F^0
			\right)
    = 2\gamma_s F^\text{inv}.
\end{equation}
Indeed, the physical current corresponds to the invariant current we found above.
These infinitesimal charges are clearly integrable and we denote the finite charges by
\begin{equation}
	Q_\lambda
		= \oint d\varphi\,
				\lambda^+ T
		= Q^\text{Vir}[\lambda^+].
\end{equation}

Under an allowed gauge transformation (that is, one satisfying \eqref{eq:sl2-ds-parameter-constr}), the connection components transform as
\begin{align}
	\delta F^0
		&= \lambda^+ \partial F^0 + F^0 \partial \lambda^+
			- \partial^{2}\lambda^+ + 2\lambda^- - 2F^- \lambda^+,\nonumber
	\\
	\delta F^-
		&= \partial \lambda^- + F^- \partial \lambda^+
			- F^- F^0 \lambda^+ + F^0\lambda^-.
\end{align}
The physical current then transforms as a Virasoro current,
\begin{align}
	\delta T
		&= \lambda^+ \partial T + 2 T \partial \lambda^+
			+ \gamma_s \partial^{3} \lambda^+.
\end{align}

\subsection*{Asymptotic symmetry algebra}
We can compute the Poisson bracket of the charges as follows,
\begin{align}
	\left\{ Q_\lambda, Q_\mu \right\}=\delta_\lambda Q_\mu
		&= -2\oint d\varphi\,
			\langle \mu, \delta_\lambda a_+ \rangle
		= \oint d\varphi\,
					\mu^+ \delta_\lambda T
        \nonumber\\
		&= \oint d\varphi\,
					 \mu^+\left(
						\lambda^+ \partial T + 2 T \partial \lambda^+
						+ \gamma_s \partial^{3} \lambda^+
					\right)
        \label{eq:vir-u1-decoupled-pb}\\
		&= Q^\text{Vir}\left[\mu^+\partial\lambda^+ - \lambda^+\partial\mu^+\right]
				+ \gamma_s \oint d\varphi\,
					  \mu^+\partial^{3}\lambda^+
          .\nonumber
\end{align}
Indeed, this is the Virasoro charge algebra.
To obtain the usual expression in terms of the Fourier modes of these charges, we can expand
\begin{equation}
	T(\varphi) = - \frac{1}{2\pi} \sum_n L_n e^{-in\varphi}.
\end{equation}
Note that we assume that the current is a $2\pi$-periodic functions.
The generators are obtained by choosing the corresponding Fourier modes as symmetry parameter,
\begin{equation}\label{eq:app-virasoro-modes}
	L_n = - \oint d\varphi\, e^{in\varphi}\, T(\varphi)
			= - Q^\text{Vir}[e^{in\varphi}].
\end{equation}
Then the Poisson bracket in \eqref{eq:vir-u1-decoupled-pb} leads to the following Poisson brackets and commutators for the Fourier modes,
\begin{align}
  \left\{ L_m, L_n \right\}
    &= -i (m-n) L_{m+n} - 2\pi i \gamma_s m^3 \delta_{m+n,0},\\
	[L_m, L_n]
		= i\left\{ L_m, L_n \right\}
		&= (m-n) L_{m+n}
			+ 2\pi\gamma_s m^3 \delta_{m+n,0},
\end{align}
Our definition of the bilinear form is related to the usual Chern--Simons level by $\gamma_s = k/4\pi$.
Recall that Einstein gravity corresponds to $k=\ell/4G$.
Therefore, the above reproduces the Brown-Henneaux central charge
\begin{equation}
	c = 24 \pi \gamma_s = 6k = \frac{3l}{2G}.
\end{equation}
To get the usual form of the Virasoro algebra, one has to shift
$L_m \to L_m + c\delta_{m,0}/24$.

\addcontentsline{toc}{section}{References}
\bibliography{NC}
\bibliographystyle{newutphys}

\end{document}